\input epsf
\input harvmac

\noblackbox

\mathchardef\varGamma="0100
\mathchardef\varDelta="0101
\mathchardef\varTheta="0102
\mathchardef\varLambda="0103
\mathchardef\varXi="0104
\mathchardef\varPi="0105
\mathchardef\varSigma="0106
\mathchardef\varUpsilon="0107
\mathchardef\varPhi="0108
\mathchardef\varPsi="0109
\mathchardef\varOmega="010A

\font\mbm = msbm10
\font\Scr=rsfs10 
\def\bb#1{\hbox{\mbm #1}}

\def\scr#1{\hbox{\Scr #1}}

\def\Mt{{\kern1em\hbox{$\tilde{\kern-1em{\scr M}}$}}}
\def\At{{\kern1em\hbox{$\tilde{\kern-1em{\scr A}}$}}}
\def\Kt{{\kern1em\hbox{$\tilde{\kern-1em{\scr K}}$}}}

%\def\Mt{\tilde{\cal M}}
%\def\At{\tilde{\cal A}}
%\def\Kt{\tilde{\cal K}}

%%%%%%%%%%%%%%%%%%%%%%%%%%%%%%%%%%%%%%%%%%

\nref\douglas{
M.~Berkooz, M.~R.~Douglas and R.~G.~Leigh,
``Branes intersecting at angles'',
Nucl.\ Phys.\ B480 (1996) 265
[arXiv:hep-th/9606139].
%%CITATION = HEP-TH 9606139;%%
}
\nref\intersectingi{
R.~Blumenhagen, L.~G\"orlich and B.~K\"ors,
``Supersymmetric orientifolds in 6D with D-branes at angles'',
Nucl.\ Phys.\ B569 (2000) 209
[arXiv:hep-th/9908130].
%%CITATION = HEP-TH 9908130;%%
}
\nref\intersectingii{
R.~Blumenhagen, L.~G\"orlich and B.~K\"ors,
``Supersymmetric 4D orientifolds of type IIA with D6-branes at angles'',
JHEP 0001 (2000) 040
[arXiv:hep-th/9912204].
%%CITATION = HEP-TH 9912204;%%
}
\nref\intersectingiii{
G.~Aldazabal, S.~Franco, L.~E.~Ib{\'a}{\~n}ez, R.~Rabadan and A.~M.~Uranga,
``$D = 4$ chiral string compactifications from intersecting branes'',
J.\ Math.\ Phys.\  42 (2001) 3103
[arXiv:hep-th/0011073],
%%CITATION = HEP-TH 0011073;%%
``Intersecting brane worlds'',
JHEP 0102 (2001) 047
[arXiv:hep-ph/0011132].
%%CITATION = HEP-PH 0011132;%%
}
\nref\intersectingiv{
R.~Blumenhagen, V.~Braun, B.~K\"ors and D.~L\"ust,
``Orientifolds of K3 and Calabi-Yau manifolds with intersecting D-branes'',
JHEP 0207 (2002) 026
[arXiv:hep-th/0206038].
%%CITATION = HEP-TH 0206038;%%
}
\nref\intersectingv{
A.~M.~Uranga,
``Local models for intersecting brane worlds'',
JHEP 0212 (2002) 058
[arXiv:hep-th/0208014].
%%CITATION = HEP-TH 0208014;%%
}
\nref\bachas{
C.~Bachas,
``A Way to break supersymmetry'',
arXiv:hep-th/9503030.
%%CITATION = HEP-TH 9503030;%%
}
\nref\us{
C.~Angelantonj, I.~Antoniadis, E.~Dudas and A.~Sagnotti,
``Type-I strings on magnetised orbifolds and brane transmutation'',
Phys.\ Lett.\ B489 (2000) 223
[arXiv:hep-th/0007090].
%%CITATION = HEP-TH 0007090;%%
}
\nref\ralphi{
R.~Blumenhagen, L.~G\"orlich, B.~K\"ors and D.~L\"ust,
``Noncommutative compactifications of type I strings on tori with  magnetic
background flux'',
JHEP 0010 (2000) 006
[arXiv:hep-th/0007024].
%%CITATION = HEP-TH 0007024;%%
}
\nref\wittenso{
E.~Witten,
``Some Properties Of O(32) Superstrings'',
Phys.\ Lett.\ B149 (1984) 351.
%%CITATION = PHLTA,B149,351;%%
}
\nref\tseytlin{
E.~S.~Fradkin and A.~A.~Tseytlin,
``Nonlinear Electrodynamics From Quantized Strings'',
Phys.\ Lett.\ B163 (1985) 123.
%%CITATION = PHLTA,B163,123;%%
}
\nref\aboob{
A.~Abouelsaood, C.~G.~.~Callan, C.~R.~Nappi and S.~A.~Yost,
``Open Strings In Background Gauge Fields'',
Nucl.\ Phys.\ B280 (1987) 599.
%%CITATION = NUPHA,B280,599;%%
}
\nref\standardi{
R.~Blumenhagen, B.~K\"ors, D.~L\"ust and T.~Ott,
``The standard model from stable intersecting brane world orbifolds'',
Nucl.\ Phys.\ B616 (2001) 3
[arXiv:hep-th/0107138].
%%CITATION = HEP-TH 0107138;%%
}
\nref\stanradii{
M.~Cveti{\v c}, G.~Shiu and A.~M.~Uranga,
``Three-family supersymmetric standard like models from intersecting  brane
worlds'',
Phys.\ Rev.\ Lett.\  87 (2001) 201801
[arXiv:hep-th/0107143],
%%CITATION = HEP-TH 0107143;%%
``Chiral four-dimensional $N = 1$ supersymmetric type IIA orientifolds from
intersecting D6-branes'',
Nucl.\ Phys.\ B615 (2001) 3
[arXiv:hep-th/0107166].
%%CITATION = HEP-TH 0107166;%%
}
\nref\standardiii{
S.~F{\"o}rste, G.~Honecker, and R.~Schreyer, 
``Supersymmetric $\bb{Z}_N \times \bb{Z}_M$ orientifolds in 4D with D-branes at angles'',
Nucl. Phys. B593  (2001) 127--154,
[arXiv:hep-th/0008250],
%%CITATION = HEP-TH 0008250;%%.
``Orientifolds with branes at angles'',  
JHEP 06 (2001) 004,
[arXiv:hep-th/0105208].
%%CITATION = HEP-TH 0105208;%%.
}
\nref\standardiv{
D.~Bailin, G.~V. Kraniotis, and A.~Love, 
``Standard-like models from intersecting D4-branes'',  
Phys. Lett. B530 (2002) 202--209,
[arXiv:hep-th/0108131].
%%CITATION = HEP-TH 0108131;%%.
}
\nref\standardv{
G.~Honecker, 
``Intersecting brane world models from D8-branes on $(T^2 \times
T^4/\bb{Z}_3 )/\varOmega R_1$ type IIA orientifolds'', 
JHEP 01 (2002) 025,
[arXiv:hep-th/0201037].
%%CITATION = HEP-TH 0201037;%%.
}
\nref\standardvi{
D.~Cremades, L.~E. Ib{\'a}{\~n}ez, and F.~Marchesano, 
``Standard model at intersecting D5-branes: Lowering the string scale'',
Nucl. Phys.  B643 (2002) 93,
[arXiv:hep-th/0205074].
%%CITATION = HEP-TH 0205074;%%.
}
\nref\standardvii{
C.~Kokorelis, 
``New standard model vacua from intersecting branes'',
JHEP 09 (2002) 029,
[arXiv:hep-th/0205147],
%%CITATION = HEP-TH 0205147;%%.
``Exact standard model compactifications from intersecting branes'', 
JHEP 08 (2002) 036,
[arXiv:hep-th/0206108],
%%CITATION = HEP-TH 0206108;%%.
``Exact standard model structures from intersecting D5- branes'',
Nucl. Phys. B677 (2004) 115,
[arXiv:hep-th/0207234].
%%CITATION = HEP-TH 0207234;%%.
}
\nref\standardviii{
D.~Bailin, G.~V. Kraniotis, and A.~Love, 
``New standard-like models from intersecting D4-branes'', 
Phys. Lett.  B547 (2002) 43,
[arXiv:hep-th/0208103],
%%CITATION = HEP-TH 0208103;%%.
``Standard-like models from intersecting D5-branes'',  
Phys. Lett. B553 (2003) 79,
[arXiv:hep-th/0210219].
%%CITATION = HEP-TH 0210219;%%.
}
\nref\standardix{
M.~Cveti{\v c}, I.~Papadimitriou, and G.~Shiu, 
``Supersymmetric three family SU(5) grand unified models from type IIA orientifolds with intersecting D6-branes'',   
Nucl. Phys. B659 (2003) 193,
[arXiv:hep-th/0212177].
%%CITATION = HEP-TH 0212177;%%.
}
\nref\standardx{
G.~Honecker, 
``Chiral supersymmetric models on an orientifold of $\bb{Z}_4 \times \bb{Z}_2$
with intersecting D6-branes'',
Nucl. Phys. B666 (2003) 175,
[arXiv:hep-th/0303015].
%%CITATION = HEP-TH 0303015;%%.
}
\nref\standardxi{
M.~Cveti{\v c} and I.~Papadimitriou, 
``More supersymmetric standard-like models from intersecting D6-branes on type IIA orientifolds'',  
Phys. Rev. D67 (2003) 126006,
[arXiv:hep-th/0303197].
%%CITATION = HEP-TH 0303197;%%.
}
\nref\standardxibis{
  M.~Axenides, E.~Floratos and C.~Kokorelis,
  %``SU(5) unified theories from intersecting branes,''
  JHEP {\bf 0310}, 006 (2003)
  [arXiv:hep-th/0307255].
  %%CITATION = HEP-TH 0307255;%%
}
\nref\standardxii{
M.~Cveti{\v c}, T.~Li, and T.~Liu, 
``Supersymmetric Pati-Salam models from intersecting D6- branes: A road to the standard model'',  
Nucl. Phys. B698 (2004) 163,
[arXiv:hep-th/0403061].
%%CITATION = HEP-TH 0403061;%%.
}
\nref\standardxiii{
G.~Honecker and T.~Ott, 
``Getting just the Supersymmetric Standard Model at Intersecting Branes on the $\bb{Z}_6$-orientifold'', 
Phys. Rev. D70  (2004) 126010,
[arXiv:hep-th/0404055].
%%CITATION = HEP-TH 0404055;%%.
}
\nref\standardxiv{
M.~Cveti{\v c}, P.~Langacker, T.-j. Li, and T.~Liu, 
``D6-brane splitting on type IIA orientifolds'',
arXiv:hep-th/0407178.
%%CITATION = HEP-TH 0407178;%%.
}
\nref\standardxv{
F.~Marchesano and G.~Shiu,
``MSSM vacua from flux compactifications'',
arXiv:hep-th/0408059,
%%CITATION = HEP-TH 0408059;%%.
``Building MSSM flux vacua'', 
JHEP 11  (2004) 041,
[arXiv:hep-th/0409132].
%%CITATION = HEP-TH 0409132;%%.
}
\nref\fbralph{
R.~Blumenhagen, B.~K\"ors and D.~L\"ust,
``Type I strings with $F$- and $B$-flux'',
JHEP 0102 (2001) 030
[arXiv:hep-th/0012156].
%%CITATION = HEP-TH 0012156;%%
}
\nref\ibanez{
L.~E.~Ib{\'a}{\~n}ez, F.~Marchesano and R.~Rabadan,
``Getting just the standard model at intersecting branes'',
JHEP 0111 (2001) 002
[arXiv:hep-th/0105155].
%%CITATION = HEP-TH 0105155;%%
}
\nref\ibwreviewi{
A.~M.~Uranga,
``Chiral four-dimensional string compactifications with intersecting D-branes'',
Class.\ Quant.\ Grav.\  20 (2003) S373
[arXiv:hep-th/0301032].
%%CITATION = HEP-TH 0301032;%%
}
\nref\ibwreviewii{  
F.~G.~Marchesano,
``Intersecting D-brane models'',
arXiv:hep-th/0307252.
%%CITATION = HEP-TH 0307252;%%
}
\nref\ibwrevirewiii{
T.~Ott,
``Aspects of stability and phenomenology in type IIA orientifolds with
intersecting D6-branes'',
Fortsch.\ Phys.\  52 (2004) 28
[arXiv:hep-th/0309107].
  %%CITATION = HEP-TH 0309107;%%
}
\nref\ibwreviewiv{
L.~G\"orlich,
``$N = 1$ and non-supersymmetric open string theories in six and four
space-time dimensions'',
arXiv:hep-th/0401040.
%%CITATION = HEP-TH 0401040;%%
}
\nref\ibwreviewv{
D.~L\"ust,
``Intersecting brane worlds: A path to the standard model?'',
Class.\ Quant.\ Grav.\  21 (2004) S1399
[arXiv:hep-th/0401156].
%%CITATION = HEP-TH 0401156;%%
}
\nref\ibwreviewvi{
R.~Blumenhagen, M.~Cveti{\v c}, P.~Langacker and G.~Shiu,
``Toward realistic intersecting D-brane models'',
arXiv:hep-th/0502005.
%%CITATION = HEP-TH 0502005;%%
}
\nref\rabadan{
R.~Rabadan,
``Branes at angles, torons, stability and supersymmetry'',
Nucl.\ Phys.\ B620 (2002) 152
[arXiv:hep-th/0107036].
%%CITATION = HEP-TH 0107036;%%
}
\nref\nielseni{
N.~K.~Nielsen and P.~Olesen,
``An Unstable Yang-Mills Field Mode'',
Nucl.\ Phys.\ B144 (1978) 376.
%%CITATION = NUPHA,B144,376;%%
}
\nref\nielsenii{
J.~Ambj\o rn, N.~K.~Nielsen and P.~Olesen,
``A Hidden Higgs Lagrangian In QCD'',
Nucl.\ Phys.\ B152 (1979) 75.
%%CITATION = NUPHA,B152,75;%%
}
\nref\nielseniii{
H.~B.~Nielsen and M.~Ninomiya,
``A Bound On Bag Constant And Nielsen-Olesen Unstable Mode In QCD'',
Nucl.\ Phys.\ B156 (1979) 1.
%%CITATION = NUPHA,B156,1;%%
}
\nref\susskind{
W.~Fischler and L.~Susskind,
``Dilaton Tadpoles, String Condensates And Scale Invariance'',
Phys.\ Lett.\ B171 (1986) 383,
%%CITATION = PHLTA,B171,383;%%
``Dilaton Tadpoles, String Condensates And Scale Invariance. 2'',
Phys.\ Lett.\ B173 (1986) 262.
%%CITATION = PHLTA,B173,262;%%
}
\nref\augvac{
E.~Dudas, G.~Pradisi, M.~Nicolosi and A.~Sagnotti,
``On tadpoles and vacuum redefinitions in string theory'',
Nucl.\ Phys.\ B708 (2005) 3
[arXiv:hep-th/0410101].
%%CITATION = HEP-TH 0410101;%%
}
\nref\scherk{
J.~Scherk and J.~H.~Schwarz,
``How To Get Masses From Extra Dimensions'',
Nucl.\ Phys.\ B153 (1979) 61,
%%CITATION = NUPHA,B153,61;%%
``Spontaneous Breaking Of Supersymmetry Through Dimensional Reduction'',
Phys.\ Lett.\ B82 (1979) 60.
%%CITATION = PHLTA,B82,60;%%
}
\nref\SSstringi{
R.~Rohm,
``Spontaneous Supersymmetry Breaking In Supersymmetric String Theories'',
Nucl.\ Phys.\ B237 (1984) 553.
%%CITATION = NUPHA,B237,553;%%
}
\nref\SSstringii{
C.~Kounnas and M.~Porrati,
``Spontaneous Supersymmetry Breaking In String Theory'',
Nucl.\ Phys.\ B310 (1988) 355.
%%CITATION = NUPHA,B310,355;%%
}
\nref\SSstringiii{
S.~Ferrara, C.~Kounnas, M.~Porrati and F.~Zwirner,
``Superstrings With Spontaneously Broken Supersymmetry And Their 
Effective Theories'',
Nucl.\ Phys.\ B318 (1989) 75;
%%CITATION = NUPHA,B318,75;%%
}
\nref\ADSi{
I.~Antoniadis, E.~Dudas and A.~Sagnotti,
``Supersymmetry breaking, open strings and M-theory'',
Nucl.\ Phys.\ B544 (1999) 469
[arXiv:hep-th/9807011].
%%CITATION = HEP-TH 9807011;%%
}
\nref\ADSii{
I.~Antoniadis, G.~D'Appollonio, E.~Dudas and A.~Sagnotti,
``Partial breaking of supersymmetry, open strings and M-theory'',
Nucl.\ Phys.\ B553 (1999) 133
[arXiv:hep-th/9812118].
%%CITATION = HEP-TH 9812118;%%
}
\nref\aldoi{
A.~L.~Cotrone,
``A $\bb{Z}_2\times  \bb{Z}_2$ orientifold with spontaneously broken supersymmetry'',
Mod.\ Phys.\ Lett.\ A14 (1999) 2487
[arXiv:hep-th/9909116].
%%CITATION = HEP-TH 9909116;%%
}
\nref\aldoii{
P.~Anastasopoulos, A.~B.~Hammou and N.~Irges,
``A class of non-supersymmetric open string vacua'',
Phys.\ Lett.\ B581 (2004) 248
[arXiv:hep-th/0310277].
%%CITATION = HEP-TH 0310277;%%
}
\nref\aldoiibis{
C.~Angelantonj and I.~Antoniadis,
``Suppressing the cosmological constant in non-supersymmetric type I strings'',
Nucl.\ Phys.\ B676 (2004) 129
[arXiv:hep-th/0307254].
%%CITATION = HEP-TH 0307254;%%
}
\lref\aldoiiter{
E.~Dudas and C.~Timirgaziu,
``Non-tachyonic Scherk-Schwarz compactifications, cosmology and moduli
stabilization,''
JHEP 0403 (2004) 060
[arXiv:hep-th/0401201].
%%CITATION = HEP-TH 0401201;%%
}
\nref\pascal{
P.~Anastasopoulos and A.~B.~Hammou,
``A classification of toroidal orientifold models'',
arXiv:hep-th/0503044.
%%CITATION = HEP-TH 0503044;%%
}
\nref\cargesei{
A.~Sagnotti,
``Open Strings And Their Symmetry Groups'',
Cargese Summer Institute on Non-Perturbative Methods in Field Theory, 
Cargese, France, 1987
[arXiv:hep-th/0208020].
%%CITATION = HEP-TH 0208020;%%
}
\nref\cargeseii{
G.~Pradisi and A.~Sagnotti,
``Open String Orbifolds'',
Phys.\ Lett.\ B216 (1989) 59.
%%CITATION = PHLTA,B216,59;%%
}
\nref\cargeseiii{
P.~Ho{\v r}ava,
``Strings On World Sheet Orbifolds'',
Nucl.\ Phys.\ B327 (1989) 461.
%%CITATION = NUPHA,B327,461;%%
}
\nref\cargeseiv{
M.~Bianchi and A.~Sagnotti,
``On The Systematics Of Open String Theories'',
Phys.\ Lett.\ B247 (1990) 517, 
%%CITATION = PHLTA,B247,517;%%
``Twist Symmetry And Open String Wilson Lines'',
Nucl.\ Phys.\ B361 (1991) 519.
%%CITATION = NUPHA,B361,519;%%
}
\nref\reviewi{
E.~Dudas,
``Theory and phenomenology of type I strings and M-theory'',
Class.\ Quant.\ Grav.\ 17 (2000) R41
[arXiv:hep-ph/0006190].
%%CITATION = HEP-PH 0006190;%%
}
\nref\reviewii{
C.~Angelantonj and A.~Sagnotti,
``Open strings'',
Phys.\ Rept.\  371 (2002) 1
[Erratum-ibid.\  376 (2003) 339]
[arXiv:hep-th/0204089].
%%CITATION = HEP-TH 0204089;%%
}
\nref\emilian{
E.~Dudas and C.~Timirgaziu,
``Internal magnetic fields and supersymmetry in orientifolds'',
arXiv:hep-th/0502085.
%%CITATION = HEP-TH 0502085;%%
}
\nref\marianna{
M.~Larosa and G.~Pradisi,
``Magnetized four-dimensional $\bb{Z}_2\times  \bb{Z}_2$ orientifolds'',
Nucl.\ Phys.\ B667 (2003) 261
[arXiv:hep-th/0305224].
%%CITATION = HEP-TH 0305224;%%
}
\nref\toroidal{
M.~Bianchi, G.~Pradisi and A.~Sagnotti,
``Toroidal compactification and symmetry breaking in open string theories'',
Nucl.\ Phys.\ B376 (1992) 365.
%%CITATION = NUPHA,B376,365;%%
}
\nref\polch{
J.~Polchinski, S.~Chaudhuri and C.~V.~Johnson,
``Notes on D-Branes'',
arXiv:hep-th/9602052.
%%CITATION = HEP-TH 9602052;%%
}
\nref\sen{
A.~Sen,
``Tachyon condensation on the brane antibrane system'',
JHEP 9808 (1998) 012
[arXiv:hep-th/9805170],
%%CITATION = HEP-TH 9805170;%%
``Non-BPS states and branes in string theory'',
arXiv:hep-th/9904207.
%%CITATION = HEP-TH 9904207;%%
}
\nref\tristan{
I.~Antoniadis and T.~Maillard,
``Moduli stabilization from magnetic fluxes in type I string theory'',
arXiv:hep-th/0412008.
%%CITATION = HEP-TH 0412008;%%
}
\nref\trevigne{
M.~Bianchi and E.~Trevigne,
``The open story of the magnetic fluxes'',
arXiv:hep-th/0502147.
%%CITATION = HEP-TH 0502147;%%
}
\nref\discrete{
C.~Angelantonj and R.~Blumenhagen,
``Discrete deformations in type I vacua'',
Phys.\ Lett.\ B473 (2000) 86
[arXiv:hep-th/9911190].
%%CITATION = HEP-TH 9911190;%%
}
\nref\augusto{
C.~Angelantonj and A.~Sagnotti,
``Type-I vacua and brane transmutation'',
arXiv:hep-th/0010279.
%%CITATION = HEP-TH 0010279;%%
}
\nref\greeni{
M.~B.~Green and J.~H.~Schwarz,
``Anomaly Cancellation In Supersymmetric $D=10$ Gauge Theory And Superstring
Theory'',
Phys.\ Lett.\ B149 (1984) 117.
%%CITATION = PHLTA,B149,117;%%
}
\nref\greenii{
A.~Sagnotti,
``A Note on the Green-Schwarz mechanism in open string theories'',
Phys.\ Lett.\ B294 (1992) 196
[arXiv:hep-th/9210127].
%%CITATION = HEP-TH 9210127;%%
}

%%%%%%%%%%%%%%%%%%%%%%%%%%%%%%%%%%%%%%%%%%%%

\Title{\vbox{
\rightline{\tt hep-th/0503179} 
\rightline{LMU-ASC 21/05}
\rightline{DFTT 07/05}
\rightline{IFUM-829-FT}
}}
{\vbox{
\centerline{Scherk-Schwarz Breaking and Intersecting Branes}
}}

\centerline{Carlo Angelantonj$^{\dagger\ast}$, Matteo Cardella$^\ddagger$ and Nikos Irges$^\star$}
\medskip
\centerline{\it $^\dagger$Arnold-Sommerfeld-Center for Theoretical Physics}
\centerline{\it Department f\"ur Physik, Ludwig-Maximilians-Universit\"at M\"unchen,}
\centerline{\it Theresienstr 37, 80333 Munich, Germany}
\centerline{\it $^\ast$ Dipartimento di Fisica Teorica, Universit\`a di Torino,}
\centerline{\it Via P. Giuria 1, 10125 Torino, Italy}
\centerline{\it $^\ddagger$ Dipartimento di Fisica, Universit\`a di Milano,}
\centerline{\it INFN sezione di Milano, via Celoria 16, 20133 Milano, Italy}
\centerline{\it $^\star$ High Energy and Elementary Particle Physics Division,}
\centerline{\it Department of Physics, University of Crete, 71003 Heraklion, Greece}

\vskip 0.3in

\centerline{\bf Abstract}

\noindent
We study the effect of Scherk-Schwarz deformations on intersecting branes.  Non-chiral fermions in {\it any} representation of the Chan-Paton gauge group generically acquire a tree-level mass dependent on the compactification radius and the  brane wrapping numbers. This offers an elegant solution to one of the long-standing problems in intersecting-brane-world models where the ubiquitous presence of massless non-chiral fermions is a clear embarrassment for any attempt to describe the Standard Model of Particle Physics. 

\Date{March, 2005}

\newsec{Introduction and conclusions}

One of the most challenging goals of String Theory is the search for four-dimensional vacua close to the observed world of Particle Physics and Einstein gravity. To this end we have witnessed in recent years enormous efforts to investigate various approaches that might lead to four-dimensional chiral spectra and eventually diverse mechanisms for breaking supersymmetry. One of the most successful scenarios so far explored is that of intersecting brane worlds \douglas\ in orientifold constructions \refs{\intersectingi {--} \intersectingv}, or its T-dual description in terms of magnetised backgrounds \refs{\bachas,\us,\ralphi}. It can be considered as a string extension of the Landau problem of quantum particle dynamics in the presence of constant magnetic fields. The Pauli couplings between the Maxwell field and 
charged particles of different helicities, together with the tower of Landau excitations offers a natural way to get chiral fermions in lower dimensions in a relatively simple setting \wittenso. One of the amazing properties of this kind of compactifications when extended to string theory is that Abelian magnetic backgrounds only affect the boundary conditions for open-strings and yield a notable class of exactly solvable conformal field theories. In turn, this allows a consistent full-fledged perturbative string theory analysis as first shown in \refs{\tseytlin,\aboob}. Aside from representing a natural setting for obtaining chiral spectra more or less close to that of the Standard Model of Particle Physics \refs{\standardi{--}\ibanez} (for reviews, see \refs{\ibwreviewi{--}\ibwreviewvi})
, magnetic fields and/or intersecting branes typically offer a valuable mechanism for breaking supersymmetry in the matter sector, a second important requirement for any string compactification to be phenomenologically appealing. Despite these many successes, magnetic fluxes alone are not however sufficient to provide a faithful string description of the Standard Model. In fact, tachyonic fields are generically present in the spectrum of light excitations, aside perhaps from some limited regions of moduli space \rabadan , reflecting the well-known Nielsen-Olesen instability \refs{\nielseni,\nielsenii,\nielseniii}, and their condensation might trigger unwanted gauge-symmetry breaking. Moreover, these non-supersymmetric vacua typically generate a sizeable vacuum energy density already at the tree level as a result of un-cancelled NS-NS tadpoles. This is a very important problem in String and Field Theory for its undesirable consequence of vacuum destabilisation, but to date no definite and successful solution has been proposed, aside from few encouraging results of old \susskind\ and recent \augvac\ investigations. Last but not the least, in intersecting brane models the spectrum of (classically) massless states always comprises some non-chiral matter, in addition to the interesting chiral excitations previously discussed. In fact, aside from non-chiral model-dependent sectors arising whenever pairs of branes are parallel in some compact dimension, the gauge vectors are {\it always} accompanied by massless fermions in the adjoint representation as well as by a number of scalars, so that some supersymmetry is locally preserved. The emergence of these additional states is quite easy to understand since they all correspond to open-strings whose ends are both attached on the same stack of (by definition parallel) branes. This is clearly an embarrassing feature of magnetic-flux compactifications since the Standard Model spectrum does not comprises such massless non-chiral fermions!

Other tools for breaking supersymmetry have however been devised both in Field Theory and in String Theory. Among these, the Scherk-Schwarz idea to relate supersymmetry breaking to compactification is very elegant and interesting \scherk. In the simplest case of circle compactification, it amounts to allowing the higher-dimensional fields to be periodic around the circle up to an R-symmetry transformation. The Kaluza-Klein momenta of the various fields are correspondingly shifted proportionally to their R charges, and modular invariance dictates the extension of this mechanism to the full perturbative spectrum in models of oriented closed strings \refs{\SSstringi,\SSstringii,\SSstringiii}. When open strings are present (in a background without fluxes), one has to distinguish between the two cases of Scherk-Schwarz deformations transverse or longitudinal to the world-volume of the branes \refs{\ADSi{--}\pascal}. In fact, in the former case the open-string fields do not depend on the coordinates of the extra dimension, and therefore are not affected by the deformation. In this scenario, termed ``M-theory breaking'', the D-brane excitations stay supersymmetric (at least to lowest order) and the fermion masses are identically vanishing, $m_{1/2} =0$. In the latter case, instead, the R charges determine the masses of the fields and $m_{1/2} \sim R^{-1}$.
It is then clear that the Scherk-Schwarz deformation can help to generate masses only in those sectors where zero-modes are present. For instance, in orbifold compactifications twisted fields living at the fixed points of the orbifold do not depend on the compact coordinates and thus stay classically supersymmetric. {\it It is then clear that this is precisely the kind of deformation needed to lift the massless non-chiral fermions while preserving intact the chiral spectrum at the brane intersections}. 

However, when Scherk-Schwarz deformations are introduced in models with branes at angle one is typically facing a new situation where the boundary conditions are altered along coordinates that are neither longitudinal to the branes nor transverse to them. Let us consider, for example, the simple situation of a brane wrapping $q$ and $k$ times  the horizontal and vertical sides of a $T^2$. Fields living on such rotated brane depend by construction on the combination $\bar y = \sqrt{ (q y_1)^2 + (k y_2)^2}$, where $y_1$ and $y_2$ label the two Cartesian coordinates of the torus of length $R_1$ and $R_2$, and their wave-function is now
$$
\phi_m (\bar y) \sim e^{2 i \pi m \bar y/ L_\|} \,,
$$
where $L_\| = \sqrt{ (q R_1)^2 + (k R_2)^2}$ is the effective length of the brane which extends along the diagonal of the multi-covering torus with sides $qR_1$ and $kR_2$. 
Then, a Scherk-Schwarz deformation $\sigma$ acting, say, along the horizontal axis of the torus is order-two in the closed-string sector but it can be trivial in the open-string sector.
In fact, enforcing Scherk-Schwarz periodicity conditions on fields on the branes implies that the Scherk-Schwarz deformation be extended to the covering space. As a result, one has to consider the action of $\sigma^q$, whose effect clearly depends on the horizontal wrapping number: {\it fields living on parallel branes with odd (even) horizontal wrapping number are (not) affected by the Scherk-Schwarz deformation, and thus the non-chiral fermions can (not) be lifted in mass}. Moreover, excitations living at brane intersections, that as such do not have zero modes, are not altered by the deformation and the chiral charged fermions stay massless. Clearly, these results can be suitably extended to the case of deformations acting along the vertical side of the torus or along any other direction.

The paper is organised as follows. In Section two we review the description of Scherk-Schwarz deformations in nine-dimensional open-string models as first discussed in \ADSi.
In Section three we turn to the general topic of intersecting branes. After we explain our notation and conventions we introduce in this context some simple deformations of the open-string sector that play an important role in the description of the Scherk-Schwarz deformations in terms of freely acting orbifolds. Section four represents the main part of this paper and describes intersecting brane models with modified boundary conditions {\it \`a la} Scherk-Schwarz. Here we follow different paths to justify our results and present toroidal  examples of Standard-Model-like spectra without massless non-chiral fermions. 
Finally, in Section five we extend our results to the case of orbifold compactifications.

\newsec{Glimpses of Scherk-Schwarz and M-theory breaking}

In this Section we quickly review the stringy realisation of Scherk-Schwarz supersymmetry breaking. At times, it can be conveniently described in terms of freely acting orbifolds where the space-time fermionic index $(-1)^F$ is combined with shifts along the compact directions \refs{\SSstringi,\SSstringii,\SSstringiii}. In field theory, where only Kaluza-Klein excitations are present, the Scherk-Schwarz mechanism can only result from shifts of internal KK momenta. On the other hand, String Theory presents more possibilities, since one has the option of affecting either momenta or windings. For oriented closed superstrings these two deformations, related by T-duality, describe essentially the same physics. However, in orientifold models \refs{\cargesei{--}\cargeseiv}
(see \refs{\reviewi,\reviewii} for reviews on orientifold constructions) different results are expected \ADSi. In the following we shall review the main features of type IIB orientifolds with supersymmetry broken via momentum or winding deformations. Actually, to emphasise the geometry of these orientifolds, we focus on the more conventional momentum-deformed closed-string spectrum and study the $\varOmega$ and $\varOmega {\scr I}$ orientifolds, where $\varOmega$ is the world-sheet parity and ${\scr I}$ is an inversion of the compact coordinate. The latter orientifold better described within type IIA is the T-dual description of type IIB with winding shifts.  

The deformed nine-dimensional spectrum of the closed IIA and IIB oriented strings is encoded in the one-loop partition function
\eqn\niness{
\eqalign{
{\scr T} =& {\textstyle{1\over 2}} \biggl[ |V_8 - S_8 |^2 \, \varLambda_{m,n} + |V_8 + S_8 |^2\, (-1)^m \, \varLambda_{m,n} 
\cr
& + |O_8 - C_8 |^2 \, \varLambda_{m,n+{1\over 2}} + |O_8 + C_8 |^2 \, (-1)^m \,\varLambda_{m,n+{1\over 2}}\biggr] \,,
\cr}
}
where $\varLambda_{m,n}$ is a $(1,1)$-dimensional Narain lattice. Indeed, all the fermions have acquired a mass proportional to the inverse radius, and a twisted tachyon is present if $R<2 \sqrt{\alpha '}$. 

The Klein bottle amplitudes associated to the two orientifolds $\varOmega$ and $\varOmega ' = \varOmega {\scr I}$ can be straightforwardly determined from \niness\ and read\footnote{$^\dagger$}{As recently shown in \emilian\ one has the additional option of symmetrising the R-R sector while acting simultaneously on the tower of Kaluza-Klein states with order-two shifts.}
$$
{\scr K} = {\textstyle{1\over 2}} \, (V_8 - S_8 ) \, P_{2m} \,,
$$
and
$$
{\scr K} \,\,' = {\textstyle{1\over 2}} \, (V_8 - S_8) \, W_n + {\textstyle{1\over 2}} \, (O_8 - C_8 ) \, W_{n+{1\over 2}} \,.
$$
The former amplitude clearly spells out the presence of O9 planes while the latter involves conjugate pairs of O8 and $\overline{{\rm O}8}$ planes sitting at the two edges of the segment $S^1 / {\scr I}$. 

More interesting is the open-string spectrum associated to these orientifolds. In the first case, it consists of open string stretched between D9 branes. In particular, these branes wrap the compact direction and thus their excitations are affected by the Scherk-Schwarz deformation. Indeed, the corresponding annulus and M\"obius-strip amplitudes
$$
{\scr A} = {\textstyle{1\over 2}} \, N^2\, \left( V_8 \, P_{2m} - S_8 \, P_{2m+1} \right) \,,
$$
and 
$$
{\scr M} = - {\textstyle{1\over 2}} \, N \, \left( \hat V_8 \, P_{2m} - \hat S_8 \, P_{2m+1} \right) \,,
$$
clearly reveal that the fermions are now massive and supersymmetry is spontaneously broken. In the second case, the D8 branes are transverse to the compact direction along which the Scherk-Schwarz deformation takes place, and thus the open-string spectrum is unaffected. However, the brane configuration has to respect the symmetries we are gauging, and therefore the pairs of image (anti-)branes have to be displaced at points diametrically opposite. In equations
$$
{\scr A}\,\, ' = \left[ {\textstyle{1\over 2}} \left( N^2 + M^2 \right) \, (V_8 - S_8)  + M\, N \, (O_8 - C_8) \right] \left( W_n
+ W_{n+{1\over 2}} \right) \,,
$$
and
$$
{\scr M} \,\, '= -{\textstyle{1\over 2}} \, (N+M) \, \hat V _8 \, \left( W_n + W_{n+{1\over 2}} \right) 
+ {\textstyle{1\over 2}}\, (N-M) \, \hat S_8 \, \left( W_n - W_{n+{1\over 2}} \right) \,.
$$
As expected the fermions are still massless (at tree level), while supersymmetry is broken only by the simultaneous presence of branes and anti-branes. 

In conclusion, we can summarise our results as follows \ADSi: if the Scherk-Schwarz deformation connects points on the brane then supersymmetry is broken in the open-string sector, otherwise image branes have to be properly added in order to respect the gauged symmetry and the fermions stay massless. 
 
\newsec{Wilson lines on magnetised branes}

In this section we review some known facts about magnetised (or intersecting) branes \refs{\intersectingi{--}\intersectingv} and comment on the role of Wilson lines and/or brane displacements\footnote{$^\ddagger$}{See \standardxiv\ for a detailed analysis of Higgsing in intersecting brane vacua, and \marianna\ for related issues.}. This will give us the opportunity to present our notation and introduce some simple deformations that however will play an important role in the forthcoming sections. 

\subsec{Preliminaries on intersecting branes: notation and conventions}

Let us focus our attention on the simple case of eight-dimensional reductions with D8 branes extending along one generic direction of the $T^2$. Lower-dimensional reductions are quite simple to study, and will be discussed in the following sections. For simplicity, we take the torus to be rectangular with horizontal and vertical sides of length $R_1$ and $R_2$, respectively. Each set of parallel branes is then identified by the pair of integers $(q, k)$ that correspond to the number of times the branes wrap the horizontal and vertical sides of the $T^2$, the two canonical one-cycles. In turn, these numbers determine the oriented angle $\phi$ between the horizontal axis of the torus and the brane itself via the relation
\eqn\dirac{
\tan \phi = {k R_2 \over q R_1} \,,
}
and it is positive (negative) if starting from the horizontal axis we move counter-clock-wise
(clock-wise) towards the brane. By simple trigonometry arguments, we can then determine the effective length of the brane
$$
L_\| = \sqrt{(q R_1)^2 + (k R_2)^2}\,,
$$
and the distance between consecutive wrappings on the $T^2$
$$
L_\perp = {R_1 R_2 \over L_\|} \,.
$$
Using the quantisation condition \dirac, $L_\|$ and $L_\perp$ can be conveniently written as
$$
L_\| = {q R_1 \over \cos\phi} = {kR_2 \over \sin\phi}\,, \qquad \quad 
L_\perp = {R_2 \over q} \cos\phi = {R_1 \over k} \sin\phi \,.
$$

Given these two quantities it is then easy to determine the zero-mode spectrum of open strings stretched on this rotated brane \refs{\bachas,\intersectingi{--}\intersectingv}
$$
M^2_{\rm z.m.} = \left( {m \over L_\|}\right)^2 + {1\over \alpha'{}^2} \left( n L_\perp\right)^2 \,.
$$
It simply consists of Kaluza-Klein states along the compact direction of the brane and of winding states between two portions of the brane in the elementary cell of the $T^2$, as indicated in figure 1.

\vbox{
\vskip 10pt
\epsfxsize 2truein
\centerline{\epsffile{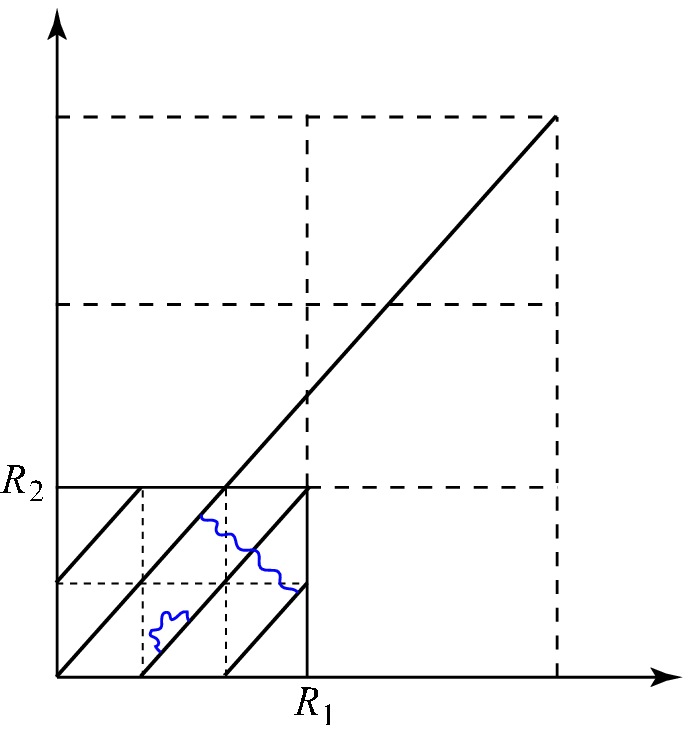}}
\vskip 10pt
\centerline{\ninepoint {\bf Figure 1.} Zero-mode spectrum on a rotated brane with wrapping numbers $(2,3)$.}
\vskip 10pt
}

If orientifold planes and/or other branes are present, one has to consider new sectors corresponding to open-strings stretched between two different branes. These new sectors are particularly appealing since they typically support chiral matter at the intersection loci, while the masses of the string excitations now depend on the relative angle $\phi_\alpha - \phi_\beta$, and the precise dependence changes in the NS and R sectors. 

As usual, it is convenient to summarise the complete spectrum of string excitations in the annulus, and eventually M\"obius-strip, partition function. To this end let us consider the simple case of an $\varOmega {\scr R}$ orientifold, with  $\varOmega$ the standard world-sheet parity and ${\scr R}:\ z\to \bar z$ an anti-conformal involution acting on the complex coordinate $z\equiv y_1 + i y_2$ of the $T^2$. This operation, a symmetry of the IIA string, introduces two pairs of horizontal O8 planes, passing through the points $y_2 =0$ and $y_2 = {1\over 2} R_2$. Introducing a stuck of $N_\alpha$ coincident D8 branes with wrapping numbers $(q_\alpha,k_\alpha)$ breaks in general the orientifold symmetry, unless a suitable stack of $N_\alpha$ image branes with wrapping numbers $(q_\alpha,-k_\alpha)$ is also added \refs{\intersectingi{--}\intersectingv}. The annulus amplitude then consists of different sectors corresponding to strings stretched between a pair of (image-)branes and to strings stretched between a brane and its image:
$$
\eqalign{
{\scr A} =& N_\alpha \bar N_\alpha \, (V_8 - S_8) [{\textstyle{0\atop 0}}]\, P_m (L_{\|\alpha}) W_n (L_{\perp\alpha})
\cr 
& + {\textstyle{1\over 2}} \left( N^2_\alpha\, (V_8 - S_8 ) [{\textstyle{\alpha\bar\alpha\atop 0}}] +  \bar N ^2_\alpha \, (V_8 - S_8) [{\textstyle{\bar\alpha \alpha\atop 0}}]
\right) \, {2 q_\alpha k_\alpha \over \varUpsilon_1 [{\textstyle{\alpha\bar\alpha\atop 0}}]}
\,.
\cr}
$$

\vbox{
\vskip 10pt
\epsfxsize 2truein
\centerline{\epsffile{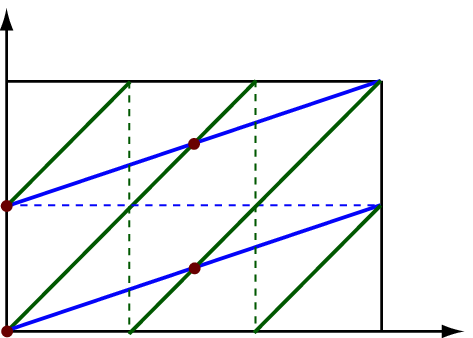}}
\vskip 10pt
\noindent
{\ninepoint {\bf Figure 2.} Two branes with wrapping numbers (2,1) and (3,2) intersecting four times on a $T^2$.}
\vskip 10pt
}
\noindent
Here 
$$
P_m (L_{\|\alpha}) = \eta^{-1}\, \sum_m q^{{\alpha ' \over 2} (m/L_{\|\alpha} )^2} \quad {\rm and}\qquad
W_n (L_{\perp\alpha} ) = \eta^{-1} \sum_n q^{{1\over 2\alpha'} (nL_{\perp\alpha} )^2 } 
$$
denote the momentum and winding lattice sums, while we have endowed the SO(8) characters with additional labels, reflecting the mass-shift of the string oscillators induced by the non-trivial intersection angle. In particular, for this $T^2$ reduction
$$
(V_8 -S_8) [{\textstyle{\alpha\beta \atop \gamma\delta}}] = V_6 \, O_2 (\zeta ) + O_6 \, V_2 (\zeta ) -  S_6 \, S_2 (\zeta) - C_6\, C_2 (\zeta)\,,
$$
with 
$$
\zeta = {1\over \pi} \left[ (\phi_\alpha - \phi_\beta ) {i\tau_2 \over 2} + \phi_\gamma - \phi_\delta \right]\,,
$$
$\tau_2$ being the proper time of the annulus and of the M\"obius-strip, and we use a barred index to define the angle $\phi_{\bar\alpha} = - \phi_\alpha$ of the image brane, whose wrapping numbers are $(q_{\bar\alpha}, k_{\bar\alpha})=(q_\alpha , - k_\alpha )$. The SO(2) characters are defined by the usual relations \refs{\reviewi,\reviewii}
$$
\eqalign{
O_2 (\zeta ) &= {e^{2i\pi \zeta} \over 2\eta} \, \left[ \vartheta_3 (\zeta |\tau) + \vartheta_4 (\zeta |\tau) \right]\,,
\cr
V_2 (\zeta ) &= {e^{2i\pi \zeta} \over 2\eta} \, \left[ \vartheta_3 (\zeta |\tau) - \vartheta_4 (\zeta |\tau) \right]\,,
\cr}\qquad
\eqalign{
S_2 (\zeta ) &= {e^{2i\pi \zeta} \over 2\eta} \, \left[ \vartheta_2 (\zeta |\tau) +i \vartheta_1 (\zeta |\tau) \right]\,,
\cr
C_2 (\zeta ) &= {e^{2i\pi \zeta} \over 2\eta} \, \left[ \vartheta_2 (\zeta |\tau) -i \vartheta_1 (\zeta |\tau) \right]\,,
\cr}
$$
with the argument of the theta functions now depending on the relative angle, and $\tau$ the Teichm\"uller parameter of the double covering torus, {\it i.e.} $\tau = i\tau_2 /2$ for the annulus amplitude and $\tau = (1+i\tau_2)/2$ for the M\"obius-strip amplitude. Finally,
$$
\varUpsilon_1 [{\textstyle{\alpha\beta \atop \gamma\delta}}] = {e^{2i\pi \zeta}\, \vartheta_1 (\zeta |\tau) \over i \eta}\,,
$$
encodes the contribution of the rotated world-sheet bosonic coordinates. Notice the multiplicative factor $2 q_\alpha k_\alpha$ in the unoriented sector. It has a simple geometrical interpretation in terms of the number of times a brane and its image intersect on the $T^2$. More generally, two branes with wrapping numbers $(q_\alpha , k_\alpha)$ and $(q_\beta , k_\beta)$ intersect a number of times given by (see fig. 2) 
$$
I_{\alpha\beta} = q_\beta k_\alpha - q_\alpha k_\beta  \,,
$$
its sign determining the chirality of massless fermions living at the intersection points \refs{\intersectingi{--}\intersectingv}.

\vbox{
\vskip 10pt
\epsfxsize 2truein
\centerline{\epsffile{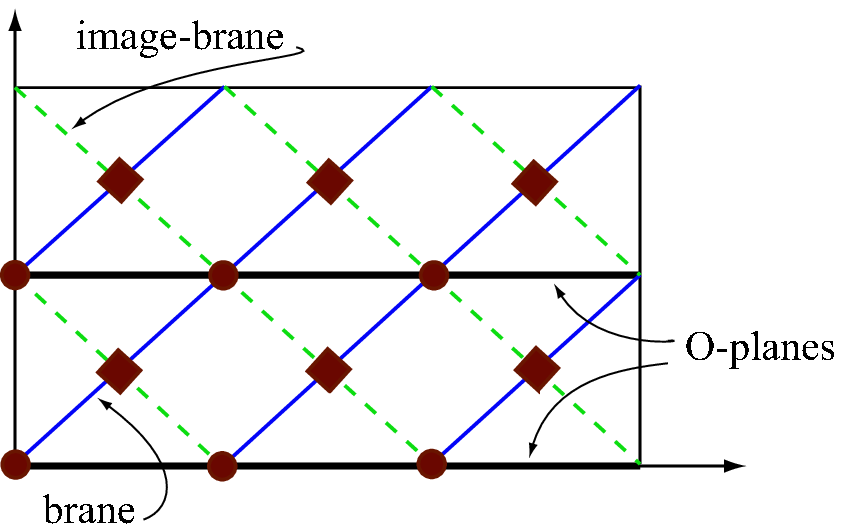}}
\vskip 10pt
\noindent
{\ninepoint {\bf Figure 3.} A brane (solid line) and its image (dashed line). The intersections denoted with a dot lie on the orientifold planes while those denoted with a diamond do not, and thus do not contribute to the M\"obius amplitude.}
\vskip 10pt
}

To conclude this brief review on intersecting branes, the M\"obius-strip amplitude
$$
{\scr M} = - {\textstyle{1\over 2}} \left( N_\alpha \, (\hat V_8 - \hat S_8) [{\textstyle{\alpha\bar\alpha\atop 0}}] + \bar N_\alpha \, (\hat V_8 - \hat S_8) [{\textstyle{\bar\alpha \alpha\atop 0}}] \right) \, {K_\alpha \over \hat\varUpsilon_1 
[{\textstyle{\alpha\bar\alpha\atop 0}}] }
$$
receives contributions only from those $K_\alpha = 2 k_\alpha$ intersections that live on the orientifold planes (see fig. 3). As a result, the massless spectrum will in general contain both symmetric and antisymmetric representations of the unitary ${\rm U} (N_\alpha)$ gauge group\footnote{$^\star$}{We postpone a detailed description of the massless spectrum to the next sections, when more physical lower-dimensional compactifications will be discussed.}.

\subsec{Wilson lines and brane displacements}

As in the more conventional case of space-filling or point-like branes, one has the option of deforming the previous spectrum by introducing suitable Wilson lines \toroidal\ or by displacing the branes in the space transverse to their world volume \polch. This simply amounts to the deformed zero-mode mass spectrum 
$$
M^2_{\rm z.m.} (a,c) = \left( {m\over L_\|} + a \right)^2 + {1\over \alpha'{}^2} (w L_\perp + c)^2 \,.
$$
It is then clear that for arbitrary values of $a$ and $c$ the open-string excitations are massive, while massless states emerge if
$$
a = {\mu \over L_\|} \,, \qquad c = \nu L_\perp\,,
$$ 
for $\mu$ and $\nu$ integers.
The latter condition simply reflects a symmetry under a rigid translation of the brane by an integer multiple of the distance between two consecutive wrappings. 

The corresponding modifications of the annulus partition function are then quite natural. Considering for simplicity the case of an orthogonal displacement $c=\delta L_\perp$ of $M_\alpha$ branes, that still have the same wrapping numbers, one has
$$
\eqalign{
{\scr A} =& (V_8 - S_8 ) [{\textstyle{0\atop 0}}] \,P_{m} (L_\|) \left[ (N_\alpha \bar N_\alpha + M_\alpha \bar M_\alpha ) W_n (L_\perp)
\right.
\cr 
&\left. +  N_\alpha \bar M_\alpha W_{n-\delta} (L_\perp ) + \bar N_\alpha M_\alpha W_{n+\delta} (L_\perp) \right]
\cr
&+ {\textstyle{1\over 2}} \biggl[ \left( N_\alpha^2 + M_\alpha^2 + 2 N_\alpha M_\alpha \right) (V_8 - S_8 ) [{\textstyle{\alpha\bar\alpha\atop 0}}] 
\cr
& + \left( \bar N_\alpha^2 + \bar M_\alpha^2 + 2 \bar N_\alpha \bar M_\alpha \right) (V_8 - S_8 ) [{\textstyle{\bar\alpha \alpha \atop 0}}] \biggr] \, {2 q_\alpha k_\alpha  \over \varUpsilon_1 [{\textstyle{\alpha\bar\alpha\atop 0}}] } \,,
\cr}
$$
with an obvious deformation of the M\"obius-strip amplitude.
It is then clear that for arbitrary $\delta$ the original gauge group ${\rm U} (N_\alpha + M_\alpha)$ is broken to ${\rm U} (N_\alpha ) \times {\rm U} (M_\alpha)$ while the (anti-)symmetric representations decompose into the sum of (anti-)symmetric representations of each group factor plus additional bi-fundamentals. As expected, for $\delta$ integer new massless vectors emerge from strings stretched between overlapping wrappings of the $N_\alpha$ and $M_\alpha$ branes and the gauge symmetry is consequently enhanced to the original ${\rm U} (N_\alpha + M_\alpha)$.

\subsec{Branes vs antibranes: an intriguing $\pi\!$uzzle}

The formalism of intersecting branes here reviewed may hide some ambiguities. It is well known in fact that anti-branes (with positive tension and negative R-R charge) are nothing but regular branes (with positive tension and positive R-R charge) that have undergone a $\pi$ rotation, as shown in fig. 4. On a compact space, if a brane has wrapping numbers $(q,k)$ and angle $\phi$ its conjugate partner has opposite wrapping numbers $(-q , -k)$ and an angle $\phi +\pi$. Despite branes and anti-branes satisfy the same quantisation condition \dirac\ and induce similar mass shifts in the spectrum of string excitations, their relative $\pi$ angle plays a crucial role in selecting the correct GSO projection. 

\vbox{
\vskip 10pt
\epsfxsize 2truein
\centerline{\epsffile{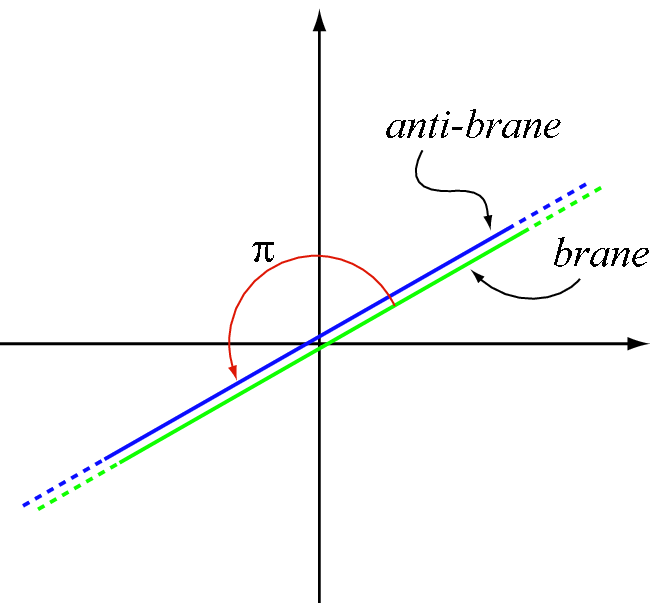}}
\vskip 10pt
\centerline{\ninepoint {\bf Figure 4.} A brane and its anti-brane differing by a $\pi$ rotation.}
\vskip 10pt
}

Let us consider in fact the spectrum of open strings stretched between two branes with angles $\phi_\alpha$ and $\phi_\beta$. As already observed, it is encoded in the partition function
$$
{(V_8 - S_8) [{\textstyle{\alpha\beta\atop 0}}] \over \varUpsilon_1 [{\textstyle{\alpha\beta \atop 0}}]} \sim {V_6 O_2 (\zeta ) + O_6 V_2  (\zeta ) -S_6 S_2 (\zeta ) - C_6 C_2  (\zeta)
\over \vartheta_1  (\zeta |\tau) \, q^{{1\over \pi}(\phi_\alpha - \phi_\beta)}} \,,
$$
with $\zeta = (\phi_\alpha - \phi_\beta)\tau/\pi$, where we have omitted irrelevant numerical factors.  After rotating the $\alpha$ brane, say, by an angle $\pi$ and converting it to an anti-brane, the partition function obviously reads
\eqn\annbab{
{(V_8 - S_8) [{\textstyle{\alpha'\beta\atop 0}}] \over \varUpsilon_1 [{\textstyle{\alpha'\beta \atop 0}}]} \sim {V_6 O_2 (\zeta_\pi ) + O_6 V_2  (\zeta_\pi ) -S_6 S_2 (\zeta_\pi ) - C_6 C_2  (\zeta_\pi) \over \vartheta_1  (\zeta_\pi  |\tau) \, q^{{1\over \pi} (\phi_\alpha - \phi_\beta) +1}} \,,
}
with now $\zeta_\pi = (\phi_\alpha - \phi_\beta+\pi)\tau/\pi$. From the very definition of theta functions
$$
\theta [{\textstyle{a\atop b}}] (z|\tau) = \sum_n q^{{1\over 2} (n + a)^2} \, e^{2i \pi (n+a)(z+b)}
\,,
$$
one can then deduce the periodicity properties
$$
\eqalign{
\vartheta_3 (\zeta_\pi |\tau ) =& + q^{- {\phi\over \pi}-{1\over 2}} \, \vartheta_3 (\zeta |\tau) \,,
\cr
\vartheta_4 (\zeta_\pi |\tau) =& - q^{- {\phi\over \pi}-{1\over 2}} \, \vartheta_4 (\zeta |\tau)\,,
\cr}
\qquad
\eqalign{
\vartheta_2 (\zeta_\pi |\tau) =& + q^{- {\phi\over \pi}-{1\over 2}} \, \vartheta_2 (\zeta |\tau )\,,
\cr
\vartheta_1 (\zeta_\pi |\tau) =& - q^{- {\phi\over \pi}-{1\over 2}} \, \vartheta_1 (\zeta |\tau)\,,
\cr}
$$
that, in turn,  induce a non-trivial reshuffling of the SO(2) characters
$$
\eqalign{
O_2 (\zeta_\pi ) =& V_2 (\zeta) \, q^{-{\phi\over \pi}+{1\over 2}}\,,
\cr 
V_2 (\zeta_\pi ) =& O_2 (\zeta) \, q^{-{\phi\over \pi}+{1\over 2}}\,,
\cr}
\qquad
\eqalign{
S_2 (\zeta_\pi ) =& C_2 (\zeta) \, q^{-{\phi\over \pi}+{1\over 2}}\,,
\cr 
C_2 (\zeta_\pi ) =& S_2 (\zeta) \, q^{-{\phi\over \pi}+{1\over 2}}\,.
\cr}
$$
As a result, the GSO projection in \annbab\ changes to
$$
{(V_8 - S_8) [{\textstyle{\alpha'\beta\atop 0}}] \over \varUpsilon_1 [{\textstyle{\alpha'\beta \atop 0}}]} \sim {V_6 V_2 (\zeta ) + O_6 O_2  (\zeta ) -S_6 C_2 (\zeta ) - C_6 S_2  (\zeta) \over \vartheta_1  (\zeta |\tau) \, q^{{1\over \pi} (\phi_\alpha - \phi_\beta) }} \,,
$$
that indeed pertains to open strings stretched between pairs of branes and anti-branes \sen.

It is evident, that the formalism introduced in sub-sections 3.1 and 3.2 is well tailored to study any kind of brane if care is used in dealing with apparently innocuous $\pi$ angles. To avoid ambiguities,  in this paper we adopt the convention to use the term brane when $q$ is positive, and anti-brane when $q$ is negative. This is clearly suggested by their contribution to the R-R tadpoles. Then according to the sign of their vertical wrapping number $k$, they can induce positive or negative charges for lower-degree gauge potentials, and consequently transmute into lower-dimensional branes (for $k>0$) or anti-branes (for $k<0$). 

\newsec{Freely acting orbifolds, Scherk-Schwarz deformations and intersecting branes}

We can now turn to the study of the effects freely acting shifts have on rotated and intersecting branes. For simplicity we shall focus to the case of a $\bb{Z}_2$ shift, although similar considerations can be straightforwardly extended to the more general case. 

On a given $T^2$ one has in principle to distinguish among the three cases of shifts acting along the horizontal axis only, along the vertical axis only or along both the horizontal and vertical axis. It turns out that it is enough to consider one case, since the others can be unambiguously determined. To this end, let us consider the shift
$$
\delta :\quad  \left\{
\eqalign{
y_1 &\to y_1 + {\textstyle{1\over 2}} R_1 \,,
\cr
y_2 & \to y_2\,,
\cr}\right.
$$
where $(y_1,y_2)$ are the natural coordinates on a rectangular torus with sides of length $R_1$ and $R_2$. With respect to the natural reference frame adapted to the brane this $\bb{Z}_2$ shift decomposes as
$$
\delta :\quad \left\{
\eqalign{
\bar y_1 &\to \bar y_1 + {\textstyle{1\over 2}} R_1 \cos\phi \,,
\cr
\bar y _2 &\to \bar y _2 - {\textstyle{1\over 2}} R_1 \sin\phi \,,
\cr} \right.
$$
with $\bar y_1$ and $\bar y _2$ labelling the directions longitudinal and transverse to the brane, respectively. This shift maps in general points on the branes to points in the bulk unless ($\mu\in\bb{Z}$)
$$
y_2 + \mu R_2 = \tan\phi \left( y_1 + {\textstyle{1\over 2}} R_1 \right) \quad \Rightarrow
\quad
\mu R_2 = {k R_2 \over 2 q} \quad \Rightarrow \quad k\in 2 \bb{Z}\,,
$$
that is to say that the shifted point still belongs to the line identified by the brane, modulo ${\rm SL} (2,\bb{Z})$ identifications of the lattice. 

\vbox{
\vskip 10pt
%\centerline{\includegraphics[width=2in]{zero-mode.pdf}}
\epsfxsize 2truein
\centerline{\epsffile{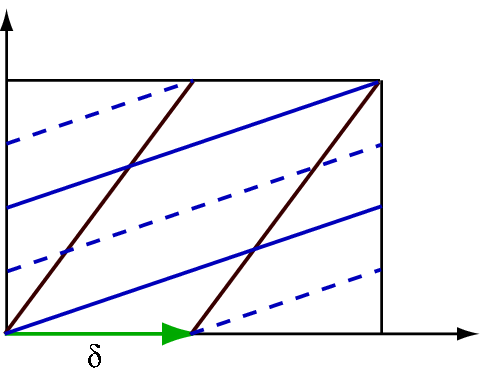}}
\vskip 10pt
\noindent
{\ninepoint {\bf Figure 5.} The two classes of branes in the presence of a freely acting horizontal shift $\delta$. The dashed brane is the image under $\delta$.}
\vskip 10pt
}

It is then evident that we can identify two equivalence classes of branes, as indicated in figure 5: those with odd $k$ and those with even $k$. In the former case the shift is not a symmetry unless we introduce image branes with the same angle but separated by a distance $- {\textstyle{1\over 2}} R_1 \sin\phi$. In the latter case the shifted points still belong to the brane and thus the given configuration is already symmetric under the action of $\delta$. Similarly, for shifts along the vertical axis (or along the diagonal of the torus) only for branes with $q$ even (or with $k$ and $q$ both odd) the image points still belong to the brane world volume.
This is somewhat reminiscent of what happens when we act with freely acting shifts along the world-volume of the brane and or along directions transverse to it.

Given these observations, it is straightforward to write down the associated vacuum amplitudes. Since in this toy model we are modding out by $\varOmega {\scr R}$, our parent theory is the type IIA superstring compactified on a $T^2$  
$$
{\scr T} = (V_8 - S_8)(\bar V _8 - \bar C_8) \, \left[ \varLambda_{2m_1,n_1} (R_1) \varLambda_{m_2,n_2} (R_2) +
\varLambda_{2m_1,n_1+{1\over 2}} (R_1) \varLambda_{m_2,n_2} (R_2)\right] \,,
$$
where each $\varLambda (R)$ denotes the Narain lattice for a circle of radius $R$ and momenta and windings specified. The associated direct-channel Klein-bottle amplitude reads
$$
{\scr K} = {\textstyle{1\over 2}} (V_8 - S_8 ) \, P_{2m_1} (R_1) W_{n_2} (R_2)  \,.
$$

\vbox{
\vskip 10pt
\epsfxsize 2truein
\centerline{\epsffile{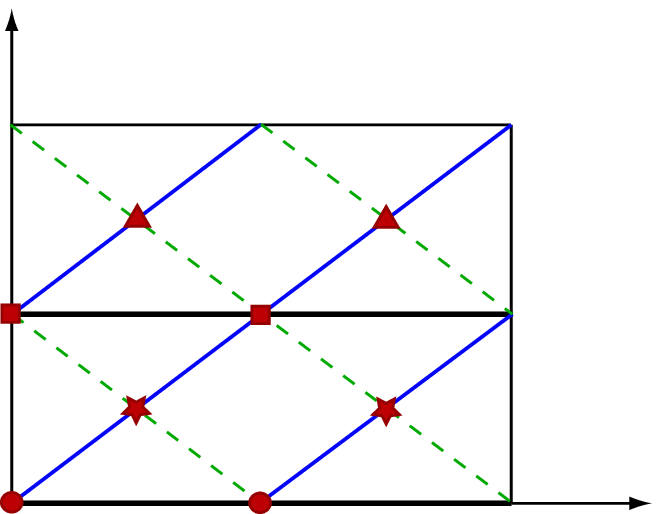}}
\vskip 10pt
\noindent
{\ninepoint {\bf Figure 6.} A (1,1) brane and its images under $\varOmega {\scr R}$ and $\delta$. The intersections marked with alike symbols are identified under $\delta$.}
\vskip 10pt
}

Moving to the open-string sector, for $k_\alpha$ odd, the direct-channel annulus and M\"obius-strip amplitudes now read
$$
\eqalign{
{\scr A} =& N_\alpha\bar N_\alpha \, (V_8 - S_8) [{\textstyle{0\atop 0}}] P_m \left[ W_{n} + W_{n+{1\over 2}} \right]
\cr
&+ {\textstyle{1\over 2}} \,
\left[ N^2_\alpha \, (V_8 - S_8) [{\textstyle{\alpha\bar\alpha \atop 0}}]
+ \bar N^2_\alpha \, (V_8 - S_8) [{\textstyle{\bar\alpha \alpha \atop 0}}]  \right] \, {2 I_{\alpha \bar\alpha} \over \varUpsilon_1 [{\textstyle{\alpha\bar\alpha \atop 0}}]} \,,
\cr}
$$ 
and
$$
{\scr M} = - {\textstyle{1\over 2}} \left[N_\alpha \, (\hat V_8 - \hat S_8) 
[{\textstyle{\alpha\bar\alpha \atop 0}}] + \bar N_\alpha \,
(\hat V_8 - \hat S_8 ) [{\textstyle{\bar \alpha \alpha \atop 0}}]
 \right] {K_\alpha \over \hat\varUpsilon_1 [{\textstyle{\alpha\bar\alpha \atop 0}}]
}
\,.
$$
Notice the important difference with the standard case: the annulus amplitude unambiguously reflects the presence of image branes under the action of the horizontal shift both by the presence of dipole strings with shifted windings, and by the doubling of the number of families for unoriented strings due to the doubling of local intersections.
As for the M\"obius-strip amplitude, instead, it is not modified since the effective number of intersections sitting on the O-planes is not affected. This counting of effective intersections is clearly depicted in figure 6 in the case of (1,1) branes, where points marked with the same symbol are identified under $\delta$. 

Moving to the case of branes with even $k_\alpha$, one is not required any longer to introduce brane images, these branes being invariant under $\delta$, and the annulus and M\"obius-strip amplitudes read
$$
\eqalign{
{\scr A} =& N_\alpha\bar N_\alpha \, (V_8 - S_8) [{\textstyle{0\atop 0}}] P_{2m} W_n
\cr
+& {\textstyle{1\over 2}} \left[ N^2_\alpha \, (V_8 -S_8) [{\textstyle{\alpha\bar\alpha \atop 0}}]
 + \bar N ^2_\alpha \, (V_8 - S_8)  [{\textstyle{\bar\alpha \alpha \atop 0}}]
\right] \, {I_{\alpha\bar\alpha} \over 2 \varUpsilon_1  [{\textstyle{\alpha\bar\alpha \atop 0}}]
 } \,,
\cr}
$$
and 
$$
{\scr M} = - {\textstyle{1\over 2}} \left[ N_\alpha \, (\hat V_8 - \hat S_8 ) 
[{\textstyle{\alpha\bar\alpha \atop 0}}]  + \bar N_\alpha  \, (\hat V _8 - \hat S_8) 
[{\textstyle{\bar \alpha \alpha \atop 0}}] \right] {K_\alpha \over 2 \hat \varUpsilon_1 
 [{\textstyle{\alpha\bar\alpha \atop 0}}] }\,.
$$
In contrast to the previous case, the multiplicities in the annulus and M\"obius-strip amplitudes have now been halved, since $\delta$ identifies pairs of intersection points, as depicted in figure 7.

\vbox{
\vskip 10pt
\epsfxsize 2truein
\centerline{\epsffile{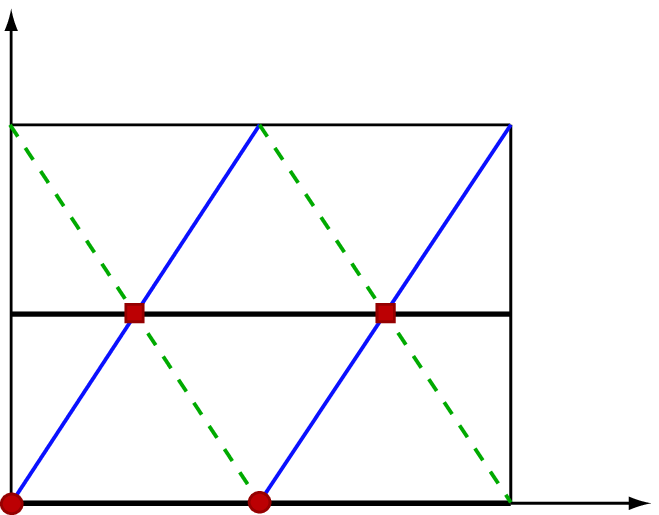}}
\vskip 10pt
\noindent
{\ninepoint {\bf Figure 7.} A (1,2) brane and its images under $\varOmega {\scr R}$. The intersections marked with alike symbols are identified under $\delta$.}
\vskip 10pt
}

We can now straightforwardly generalise the open-string amplitudes to the case of a more interesting $T^6= T^2 \times T^2\times T^2$ compactification with the shift acting along  the horizontal axis $y_1$ of the first $T^2$. For simplicity we shall focus our attention to the case of factorisable D6 branes or, in the T-dual language, to commuting magnetic backgrounds, although more general choices turned out to be very promising in stabilising closed-string moduli \refs{\tristan,\trevigne}. In this simple case, D6 branes on a $T^2 \times T^2\times T^2$ are then identified by their angles $\phi^\varLambda_\alpha$ ($\varLambda = 1,2,3$ labelling the three two-tori) and wrapping numbers $(q_\alpha^\varLambda \,,\, k^\varLambda_\alpha)$ related, as usual, by
$$
\tan \phi^\varLambda_\alpha = {k^\varLambda_\alpha \, R^\varLambda_2 \over q^\varLambda_\alpha \, R^\varLambda_1} \,,
$$
where $R^\varLambda_1$ and $R^\varLambda_2$ denote the sizes of the horizontal and vertical sides of the $\varLambda$-th $T^2$, respectively. 
We then split the generic index $\alpha$ into the pair $a$ and $i$ labelling respectively $n_{\rm o}$ different stacks of $N_a$ branes ($a=1,\ldots,n_{\rm o}$) with $k^1_a$ odd and $n_{\rm e}$ different stacks of $N_i$ branes ($i=1,\ldots,n_{\rm e}$) with $k^1_i$ even. The annulus amplitude is then
\eqn\torann{
\eqalign{
{\scr A} =& \sum_{a=1}^{n_{\rm o}} N_a \bar N_a \, (V_8 - S_8) [{\textstyle{0\atop 0}}] \, P_m^1 \left[ W_n^1 + W_{n+{1\over 2}}^1\right] P_m^2 W_n^2  P_m^3 W_n^3 
\cr
& + \sum_{i=1}^{n_{\rm e}} N_i \bar N_i \, (V_8 - S_8) [{\textstyle{0\atop 0}}] \, P_{2m}^1 W_n^1 P_m^2 W_n^2 P_m^3 W_n^3
\cr
&+ {\textstyle{1\over 2}} \sum_{a=1}^{n_{\rm o}} \left( N_a^2 (V_8 - S_8) [{\textstyle{a\bar a\atop 0}}]
+ \bar N_a^2 (V_8 - S_8)
[{\textstyle{\bar a  a \atop 0}}] \right) \, 
{2 I_{a\bar a}  \over \varUpsilon_1 [{\textstyle{a\bar a\atop 0}}]}
\cr
&+ {\textstyle{1\over 2}} \sum_{i=1}^{n_{\rm e}} \left( N_i^2 (V_8 - S_8)
[{\textstyle{i\bar\imath\atop 0}}] + \bar N_i^2 (V_8 - S_8) 
[{\textstyle{\bar\imath i\atop 0}}] \right) \, 
{I_{i\bar\imath} \over 2 \,\varUpsilon_1 [{\textstyle{i\bar\imath\atop 0}}]}
\cr
&+ \sum_{{a,b=1 \atop b<a}}^{n_{\rm o}} \left( N_a \bar N_b (V_8 - S_8 ) 
[{\textstyle{a b\atop 0}}] + \bar N_a N_b (V_8 - S_8 ) [{\textstyle{\bar a \bar b\atop 0}}]\right)\,
{ 2\, I_{a b} \over \varUpsilon_1 [{\textstyle{a b\atop 0}}]}
\cr
&+ \sum_{{a,b=1 \atop b<a}}^{n_{\rm o}} \left( N_a  N_b (V_8 - S_8 ) 
[{\textstyle{a\bar b\atop 0}}] + \bar N_a \bar N_b (V_8 - S_8 ) 
[{\textstyle{\bar a b\atop 0}}] \right) \,
{ 2\, I_{a\bar b} \over \varUpsilon_1 [{\textstyle{a\bar b\atop 0}}]}
\cr
&+ \sum_{{i,j=1 \atop j<i}}^{n_{\rm e}} \left( N_i \bar N_j (V_8 - S_8 ) 
[{\textstyle{i j\atop 0}}]
+ \bar N_i N_j (V_8 - S_8 ) [{\textstyle{\bar\imath \bar\jmath\atop 0}}]\right) \,
{ I_{ij} \over 2\, \varUpsilon_1 [{\textstyle{ij\atop 0}}]}
\cr
&+ \sum_{{i,j=1 \atop j<i}}^{n_{\rm e}} \left( N_i  N_j (V_8 - S_8 ) 
[{\textstyle{i\bar\jmath\atop 0}}] + \bar N_i \bar N_j (V_8 - S_8 ) 
[{\textstyle{\bar\imath j\atop 0}}] \right) \,
{ I_{i\bar\jmath} \over 2\, \varUpsilon_1 [{\textstyle{i\bar\jmath\atop 0}}]}
\cr
&+ \sum_{a=1}^{n_{\rm o}} \sum_{i=1}^{n_{\rm e}} \left( N_a \bar N_i \, (V_8 -S_8) 
[{\textstyle{a i\atop 0}}]+ \bar N_a N_i \, (V_8 - S_8) 
[{\textstyle{\bar a \bar\imath\atop 0}}] \right) \, 
{I_{a i} \over \varUpsilon_1 [{\textstyle{a i \atop 0}}]}
\cr
&+ \sum_{a=1}^{n_{\rm o}} \sum_{i=1}^{n_{\rm e}} \left( N_a N_i \, (V_8 -S_8) 
[{\textstyle{a\bar\imath \atop 0}}]+ \bar N_a \bar N_i \, (V_8 - S_8) 
[{\textstyle{\bar a i\atop 0}}] \right) \, {I_{a\bar\imath} \over \varUpsilon_1 [{\textstyle{a\bar\imath \atop 0}}]} \,.
\cr}
}
To lighten the notation, we omit here and in following similar expressions the ``effective radius'' dependence of the various momentum and winding lattice contributions. In particular, $P_m^\varLambda$ ($W_n^\varLambda$) is a short-hand notation for $P_m (L_{\alpha\|}^\varLambda)$ ($W_n (L_{\alpha\perp}^\varLambda)$), where $L_{\alpha\|}^\varLambda$ ($L_{\alpha\perp}^\varLambda$) is the total length (the transverse distance among the consecutive wrappings) of the $\alpha$-th stack of branes in the $\varLambda$-th torus.
The M\"obius-strip amplitude instead is a simple combination of the previous ones and reads
$$
\eqalign{
{\scr M} =& - {\textstyle{1\over 2}} \sum_{a=1}^{n_{\rm o}}
\left(N_a \, (\hat V_8 - \hat S_8) [{\textstyle{a\bar a\atop 0}}] + \bar N_a \,
(\hat V_8 - \hat S_8 ) [{\textstyle{\bar a a\atop 0}}] \right) {K_a \over \hat\varUpsilon_1 
[{\textstyle{a\bar a\atop 0}}]}
\cr
&- {\textstyle{1\over 2}} \sum_{i=1}^{n_{\rm e}}
\left( N_i\, (\hat V_8 - \hat S_8 ) [{\textstyle{i\bar\imath\atop 0}}]
+ \bar N_i \, (\hat V _8 - \hat S_8) [{\textstyle{\bar \imath i\atop 0}}] 
\right) {K_i \over 2\, \hat \varUpsilon_1 [{\textstyle{i \bar\imath\atop 0}}]}
\,.
\cr}
$$
Here we have adapted our notation to the case of multiple $T^2$'s. As already stated we append an index $\varLambda$ to angles and wrapping numbers relative to the $\varLambda$-th torus, while now
$$
I_{\alpha\beta} = \prod_{\varLambda =1}^3 \, \left( q_\beta^\varLambda \, k_\alpha^\varLambda - q_\alpha^\varLambda \, k_\beta^\varLambda \right)
$$
and
$$
K_\alpha = \prod_{\varLambda = 1,2,3} \, 2 \, k^\varLambda_\alpha
$$
count the total number of intersections in the $T^6$, and those sitting on the O6 planes. Moreover, the contribution of the world-sheet bosons is  
$$
\varUpsilon_1 \, [{\textstyle{\alpha\beta \atop \gamma\delta}}]
= \prod_{\varLambda=1,2,3} \, {\vartheta_1 (\zeta^\varLambda |\tau) \over i\eta (\tau)} \,
e^{2i \pi \zeta^\varLambda} \,,
$$ 
with
$$
\zeta^\varLambda = {1\over \pi}\left[
\left( \phi^\varLambda_\alpha - \phi^\varLambda_\beta \right) \tau + \left( \phi_\gamma^\varLambda - \phi_\delta^\varLambda \right) \right] \,.
$$
The contribution of the world-sheet fermions is more involved and requires an ${\rm SO} (8)\to {\rm SO} (2) \times {\rm SO} (2) \times {\rm SO} (2) \times {\rm SO} (2)$ breaking of the original characters. This is a consequence of the fact that, in the T-dual language of magnetised backgrounds, fields with different helicities couple differently to the magnetic fields. More explicitly
$$
\eqalign{
V_8 \, [{\textstyle{\alpha\beta \atop \gamma\delta}}] = V_2 &\, \left[
O_2 (\zeta^1 ) \, O_2 (\zeta^2) \, O_2 (\zeta^3) + 
O_2 (\zeta^1 ) \, V_2 (\zeta^2) \, V_2 (\zeta^3) \right.
\cr
&\left.+ 
V_2 (\zeta^1 ) \, V_2 (\zeta^2) \, O_2 (\zeta^3) +
V_2 (\zeta^1 ) \, O_2 (\zeta^2) \, V_2 (\zeta^3) \right]
\cr
+ O_2 &\, \left[
V_2 (\zeta^1 ) \, O_2 (\zeta^2) \, O_2 (\zeta^3) + 
V_2 (\zeta^1 ) \, V_2 (\zeta^2) \, V_2 (\zeta^3) \right.
\cr
&\left. +
O_2 (\zeta^1 ) \, V_2 (\zeta^2) \, O_2 (\zeta^3) +
O_2 (\zeta^1 ) \, O_2 (\zeta^2) \, V_2 (\zeta^3) \right] \,,
\cr
S_8 \, [{\textstyle{\alpha\beta \atop \gamma\delta}}] = S_2 &\, \left[
S_2 (\zeta^1 ) \, S_2 (\zeta^2) \, S_2 (\zeta^3) +  
S_2 (\zeta^1 ) \, C_2 (\zeta^2) \, C_2 (\zeta^3) \right.
\cr
&\left. + 
C_2 (\zeta^1 ) \, S_2 (\zeta^2) \, C_2 (\zeta^3) + 
C_2 (\zeta^1 ) \, C_2 (\zeta^2) \,S_2 (\zeta^3) \right]
\cr
+ C_2 &\, \left[
C_2 (\zeta^1 ) \, S_2 (\zeta^2) \, S_2 (\zeta^3) + 
C_2 (\zeta^1 ) \, C_2 (\zeta^2) \, C_2 (\zeta^3) \right.
\cr
&\left. +
S_2 (\zeta^1 ) \, S_2 (\zeta^2) \, C_2 (\zeta^3) +
S_2 (\zeta^1 ) \, C_2 (\zeta^2) S_2 (\zeta^3) \right] \,.
\cr}
$$
As usual, the massless spectrum can be extracted expanding ${\scr A}$ and ${\scr M}$. Aside from the full ${\scr N} = 4$ super-Yang-Mills multiplet in the adjoint representation of the Chan-Paton gauge group, one typically gets tachyonic excitations and (non-)chiral fermions in various representations. The spectrum of chiral fermions can be easily computed and is encoded in table 1, where, as usual, $A_\alpha$ and $S_\alpha$ denote respectively the anti-symmetric and symmetric representations of the $\alpha$-th factor in the gauge group. As for non-chiral fermions, these emerge whenever branes are parallel in a given $T^2$. Their multiplicity is then given by  the number of times the corresponding branes intersect in the remaining tori. For instance, if the branes of type $\alpha$ and $\beta$ are parallel in the first $T^2$, then one finds 
$$
I^{\rm non\ chiral}_{\alpha\beta} = \prod_{\varSigma=2,3} \left( q_\beta^\varSigma \, k^\varSigma_\alpha - q_\alpha^\varSigma \, k^\varSigma_\beta \right)
$$
non-chiral fermions in the representation $(N_\alpha , \bar N_\beta )$. 

\vskip 10pt
\vbox{
\settabs=6\columns
\centerline{\vbox{\hrule width  8 truecm}}
\+
&&Rep. & $\quad$multiplicity 
\cr
\centerline{\vbox{\hrule width  8 truecm}}
\+
&&$A_a$ & $\quad\quad {1\over 2} (2\, I_{a\bar a} \pm K_a)$
\cr
\+
&&$S_a$ & $\quad\quad {1\over 2} (2\, I_{a\bar a} \mp K_a)$
\cr
\+
&&$A_i$ & $\quad\quad {1\over 4} (I_{i\bar\imath} \pm K_i)$
\cr
\+
&&$S_i$ & $\quad\quad {1\over 4} (I_{i\bar\imath} \mp K_i)$
\cr
\+
&&$(N_a , N_b)$ & $\quad\quad 2\, I_{a\bar b}$
\cr
\+
&&$(N_a , \bar N_b)$ & $\quad\quad 2\, I_{a b}$
\cr
\+
&&$(N_i , N_j)$ & $\quad\quad {1\over 2}\, I_{i\bar\jmath}$
\cr
\+
&&$(N_i , \bar N_j)$ & $\quad\quad {1\over 2}\, I_{ij}$
\cr
\+
&&$(N_a , N_i)$ & $\quad\quad I_{a\bar\imath}$
\cr
\+
&&$(N_a , \bar N_i)$ & $\quad\quad I_{ai}$
\cr
\centerline{\vbox{\hrule width  8 truecm}}
\cleartabs
\vskip 10pt
\noindent
{\ninepoint {\bf Table 1.} Spectrum of four-dimensional chiral fermions. The sign of $I_{\alpha\bar\beta}$ determines the chirality of the four-dimensional spinors, while the two signs in the multiplicity of (anti-)symmetric representations refer to branes ($q>0$) or to anti-branes ($q<0$), respectively.}
\vskip 10pt
}

Turning to the transverse channel, the massless tadpoles can be extracted as usual from the leading terms in $\Kt$, $\At$ and $\Mt$ \refs{\reviewi,\reviewii}. Introducing the combinations
\eqn\radii{
{\bf R} = \prod_{\varLambda =1}^3 \, \sqrt{ R_1^\varLambda \over R_2^\varLambda}
\,,
\qquad {\bf L}_\alpha = \prod_{\varLambda = 1}^3 \, \sqrt{ L^\varLambda_{\|\, \alpha}
\over L^\varLambda_{\perp\,\alpha}}\,,
}
the NS-NS dilaton tadpole reads
\eqn\nstadpole{
2\, \sum_{a=1}^{n_{\rm o}} \, {\bf L}_a \, \left( N_a + \bar N_a \right) + \sum_{i=1}^{n_{\rm e}} \, {\bf L}_i \, \left( N_i + \bar N_i \right) = 2^5 \, {\bf R}\,.
}
Notice that, after a T-duality along the vertical axis, the left-hand-side is nothing but the Dirac-Born-Infeld Action
$$
{\bf L_\alpha} = \prod_{\varLambda=1,2,3} \, \sqrt{ R_1 R_2 \, {\rm det}\,\left[ q^\varLambda_\alpha \, \left( {\bf 1} + F^\varLambda_\alpha \right) \right]} \,,
$$
with $F^\varLambda$ a two-by-two antisymmetric matrix, whose only independent entry is the magnetic field background $H^\varLambda$. 
The eight different R-R tadpoles can be straightforwardly obtained from \nstadpole\  
after a multiplication of the Chan-Paton multiplicities by a phase depending on the scalar product of the rotation angles $\phi^\varLambda_\alpha$ and the helicities $\eta^\varLambda = \pm {1\over 2}$ of the internal spinors. With our convention such that $S_2$ has helicity $+{1\over 2}$ and $C_2$ has helicity $-{1\over 2}$, they read 
\eqn\rrtadpole{
2\, \sum_{a=1}^{n_{\rm o}} \, {\bf L}_a \, \left( N_a \, e^{2i \phi_a \cdot \eta}
+ \bar N_a \, e^{-2i \phi_a \cdot \eta} \right) + \sum_{i=1}^{n_{\rm e}} \, {\bf L}_i \, \left( N_i \, e^{2i\phi_i \cdot \eta}  + \bar N_i \, e^{-2i\phi_i \cdot \eta}\right) = 2^5 \, {\bf R}\,,
}
with the internal product $\phi_\alpha \cdot \eta = \sum_{\varLambda =1}^3 \, \phi_\alpha^\varLambda \, \eta^\varLambda$. This single R-R tadpole actually encodes couplings to different forms as dictated by the (T-dualised) Action
$$
S \sim \int \, \sum_p \, C_{p+1} \, e^{i F} \sim \int \, C_{10} + i \, C_8 \wedge F - {\textstyle{1\over 2}}\, C_6 \wedge F \wedge F - {\textstyle{1\over 6}} \,i \,C_4 \wedge F \wedge F \wedge F \,. 
$$
Each factor, in turn, receives contributions from several terms according to which internal $T^2$ the magnetic field points to.
For instance, the real part of the tadpole \rrtadpole\ yields the conditions \refs{\intersectingi{--}\intersectingv}
$$
\eqalign{
\left[
\sum_{a=1}^{n_o} \, 2 \,  q^1_a \, q^2_a \, q^3_a \, N_a +
\sum_{i=1}^{n_e} \, q^1_i \, q^2_i \, q^3_i \, N_i\right] 
\sqrt{R^1_1 \, R^2_1 \, R^3_1 \over R_2^1 \, R_2^2 \, R_2^3}=& 16 \,
\sqrt{R^1_1 \, R^2_1 \, R^3_1 \over R_2^1 \, R_2^2 \, R_2^3} \,,
\cr
\left[
\sum_{a=1}^{n_o} \, 2 \,  q^1_a \, k^2_a \, k^3_a \, N_a +
\sum_{i=1}^{n_e} \, q^1_i \, k^2_i \, k^3_i \, N_i \right] 
\sqrt{R^1_1 \, R^2_2 \, R^3_2 \over R_2^1 \, R_1^2 \, R_1^3}=& 0\,,
\cr
\left[ \sum_{a=1}^{n_o} \, 2 \,  k^1_a \, q^2_a \, k^3_a \, N_a +
\sum_{i=1}^{n_e} \, k^1_i \, q^2_i \, k^3_i \, N_i \right]
\sqrt{R^1_2 \, R^2_1 \, R^3_2 \over R_1^1 \, R_2^2 \, R_1^3}=& 0\,,
\cr
\left[
\sum_{a=1}^{n_o} \, 2 \,  k^1_a \, k^2_a \, q^3_a \, N_a +
\sum_{i=1}^{n_e} \, k^1_i \, k^2_i \, q^3_i \, N_i \right]
\sqrt{R^1_2 \, R^2_2 \, R^3_1 \over R_1^1 \, R_1^2 \, R_2^3}=& 0\,.
\cr}
$$
After the vertical axis of the three $T^2$'s are properly T dualised, one can then read the couplings with the ten-form potential $C_{10}$ and with the three six-form potentials $C^\varLambda_6$, whose six indices point to the four non-compact space-time directions and to the two compact coordinates of the $\varLambda$-th torus. The imaginary part of the tadpole \rrtadpole\ is proportional to $N_\alpha - \bar N_\alpha$ and thus vanishes identically. This is fully consistent with the fact that the $C_8$ and $C_4$ forms are projected out by the orientifold involution and thus do not belong to the physical spectrum.

\subsec{Breaking supersymmetry in the dipole-string sector}

After we have learned how shifts act on rotated branes, we can now proceed to the study of the combined effect of $\delta$ and $(-1)^F$. As is well known the (freely-acting) orbifold generated by $\delta \, (-1)^F$ is a convenient string description of Sherck-Schwarz deformations \refs{\SSstringi{--}\ADSii} that in general are responsible for giving masses to fermions, and thus for breaking supersymmetry. Following the lines of section 3 we expect that the space-time fermion index has a non trivial effect only when the shift $\delta$ moves points along the brane, that is only when $k_i$ is an even number. In this case the annulus amplitude reads
\eqn\openss{
\eqalign{
{\scr A} =& N_i \bar N_i  \, \left( V_8 [{\textstyle{0\atop 0}}]
\, P_{2m} - S_8 [{\textstyle{0\atop 0}}]
\, P_{2m+1} \right) \, W_n 
\cr
&+ {\textstyle{1\over 2}} \left( N^2_i \, (V_8 - S_8) [{\textstyle{i\bar\imath \atop 0}}]
+ \bar N ^2_i (V_8 - S_8 ) [{\textstyle{\bar\imath i\atop 0}}]
\right) \, {I_{i\bar\imath} \over 2\, \varUpsilon_1 [{\textstyle{i\bar\imath \atop 0}}]
} \,.
\cr}
}
As is spelled out by this direct-channel amplitude, the adjoint fermions have now got a mass 
$$
m_{1/2} \sim L_\|^{-1} = {\cos \phi_i \over q_i R_1} = {\sqrt{1 + \alpha ' {}^2 H_i^2} \over q_i R_1}
= {1 \over q_i^2 \, R_1^2} \sqrt{ q_i^2 \, R_1^2 + \alpha ' {}^2 {k_i^2 \over R_2^2}}\,,
$$ 
of the order of the TeV,
and, consequently, supersymmetry is broken in the neutral dipole-string sector. In the transverse channel
$$
\eqalign{
\At =& \, 2^{-6} \, {L_{\| i} \over L_{\perp i}} \, N_i \bar N_i \, \left[ ( V_8 - S_8) 
[{\textstyle{0 \atop 0}}] \,\tilde W_{n}
+ (O_8 - C_8) [{\textstyle{0\atop 0}}]
 \,\tilde W_{n+{1\over 2}} \right] \, \tilde P_m
\cr
&+ 2^{-5} \, \left[ N^2_i \, (V_8 - S_8) [{\textstyle{0 \atop i\bar\imath}}]
 + \bar N ^2_i \, (V_8 - S_8) [{\textstyle{0 \atop \bar\imath i}}]
  \right]\,
{I_{i\bar\imath} \over2\, \varUpsilon_1 [{\textstyle{0 \atop i\bar\imath}}]} \,,
\cr}
$$
the twisted closed-string R-R states are massive, and the only massless tadpoles one is to worry about are those for the untwisted R-R fields in $S_8$.

The generalisation to higher-dimensional tori is also straightforward. The only modification is in the dipole sector of the $N_i$ branes that now would read
\eqn\nonsusyann{
\sum_{i=1}^{n_{\rm e}} M_i \bar M_i \, \left( V_8 \, [{\textstyle{0\atop 0}}] \, P_{2m}^1
\, - S_8 \, [{\textstyle{0\atop 0}}] \, P_{2m+1}^1 \right)
 W_n^1 P_m^2 W_n^2 P_m^3 W_n^3 \,.
}
It reflects itself in the transverse-channel contribution
$$
2^{-6} \, 
\sum_{i=1}^{n_{\rm e}} \, ({\bf L}^\varLambda _i)^2\,  
N_i \bar N_i \, \Bigl[ ( V_8 - S_8) [{\textstyle{0\atop 0}}] 
\,\tilde W_{n}^1 
 + (O_8 - C_8) [{\textstyle{0\atop 0}}] \,\tilde W_{n+{1\over 2}}^1 \Bigr] \, \tilde P_m^1 
\tilde W_n^2 \tilde P_m^2 \tilde W_n^3 \tilde P_m^3 \,,
$$
and thus, also in this case, the massless NS-NS and R-R tadpoles \nstadpole\ and \rrtadpole\ are not affected.

\subsec{Lifting non-adjoint non-chiral fermions}

As we have remarked in previous subsections, non-chiral fermions arise not only in the dipole-string sector, where open strings in the adjoint representation end on the same stuck of branes. In fact, it may well happen that although branes intersect at non-trivial angles 
in some tori, they are actually parallel in one $T^2$. This is typically reflected by a vanishing intersection number $I_{\alpha \beta} =0$ for the given pair of $\alpha$-type and $\beta$-type branes, and in turn by an undetermined ${0\over 0}$ expression in the annulus amplitude. For concreteness, let us suppose that the $\alpha$ and $\beta$ branes are parallel in the $\varSigma$-th  torus. By definition, this implies that their angles are the same
$\phi^\varSigma_\alpha = \phi^\varSigma_\beta$ and thus also the corresponding wrapping numbers coincide
$$
q^\varSigma_\alpha \equiv q^\varSigma_\beta \,, \qquad 
k^\varSigma_\alpha \equiv k^\varSigma_\beta \,.
$$
The internal SO(2) characters corresponding to the first $T^2$ have then vanishing argument, and can be combined with the space-time SO(2) little group to reconstruct a full SO(4) symmetry, whose spinor representation is vector-like from the four-dimensional viewpoint. 
As a result, in the $N_\alpha \bar N_\beta$ sector of the annulus amplitude
$$
\eqalign{
{\scr A} =& \left( N_\alpha \bar N_\beta \, (V_8 - S_8) [{\textstyle{\alpha \beta \atop 0}}]
+ \bar N_\alpha N_\beta \, (V_8 - S_8) [{\textstyle{\bar\alpha\bar\beta \atop 0}}] \right) 
{I_{\alpha \beta} \over \varUpsilon_1  [{\textstyle{\alpha \beta \atop 0}}]}
\cr
=& \left( N_\alpha \bar N_\beta \, (V_8 - S_8) [{\textstyle{\alpha \beta \atop 0}}]
+ \bar N_\alpha N_\beta \, (V_8 - S_8) [{\textstyle{\bar\alpha\bar\beta \atop 0}}] \right) 
\prod_{\varLambda =1}^3 \, {(k^\varLambda_\alpha q^\varLambda_\beta - 
k^\varLambda_\beta q^\varLambda_\alpha )\, i\eta \, e^{2 i (\phi^\varLambda_\alpha - \phi^\varLambda_\beta)\tau} \over \vartheta_1 ({1\over \pi} (\phi^\varLambda_\alpha - \phi^\varLambda_\beta)\tau |\tau )}
\cr}
$$
the contribution from the $\varSigma$-th torus is clearly ill defined since both $k^\varSigma_\alpha q^\varSigma_\beta - k^\varSigma_\beta q^\varSigma_\alpha$ and $\vartheta_1 (0|\tau)$ vanish. Actually, one has to be very careful in cases like this, since if on the one hand it is true that parallel branes do not have any longer the tower of Landau levels, it is evident on the other hand that open strings stretched between such branes have now non-trivial zero modes, and thus their contribution has to be taken into account. In practice, this amounts to the substitution
$$
{(k^\varSigma_\alpha q^\varSigma_\beta - k^\varSigma_\beta q^\varSigma_\alpha )\, i\eta  \over \vartheta_1 (0 |\tau )} \to P_m (L^\varSigma_{\| \, \alpha} ) \, W_n (L^\varSigma_{\perp\, \alpha} ) \,.
$$
It is then clear that, acting with the $\delta (-1)^F$ deformation on the $\varSigma$-th torus,  non-chiral fermions can acquire a tree-level mass $m_{1/2} \sim 1/L_{\|\,\alpha}^\varSigma$
if their vertical wrapping number $k_\alpha^\varSigma$ is an even integer. In this case, the corresponding sector in the annulus amplitude would read
$$
N_\alpha \bar N_\beta \, \left( V_8 [{\textstyle{\alpha \beta \atop 0}}] P_{2m} (
L^\varSigma_{\| \, \alpha} ) - S_8 [{\textstyle{\alpha \beta \atop 0}}] P_{2m+1} (
L^\varSigma_{\| \, \alpha} ) \right) \, W_n (L^\varSigma_{\perp\, \alpha} ) 
\, {I^{\rm non\ chiral}_{\alpha\beta}\over \tilde\varUpsilon_1 [{\textstyle{\alpha \beta \atop 0}}]} \,,
$$
where clearly the $\varSigma$-th torus does not contribute to $I^{\rm non\ chiral}_{\alpha\beta}$ and $\tilde\varUpsilon_1$.

\subsec{Comments on Scherk-Schwarz and orbifold basis}

In the previous sub-section we have studied the effect of the combined action of the space-time fermion index and momentum shift along a compact coordinates and we have identified it with the Scherk-Schwarz mechanism \scherk. While correct in spirit, this definition does not correspond, however, to the common use of the term in Field Theory, since the canonical Scherk-Schwarz deformation for a circle would lead to periodic bosons and anti-periodic fermions, a choice manifestly compatible with any low-energy effective field theory, where fermions only enter via their bi-linears. On the other hand, from eq. \niness , rewritten more explicitly as
$$
\eqalign{
{\scr T} =& \left( |V_8 |^2 + |S_8 |^2 \right) \, \varLambda_{2m ,n} (R)- \left(S_8 \bar V_8 + V_8 \bar S_8 \right) \, \varLambda_{2m+1,n} (R)
\cr
&+ \left( |O_8|^2 + |C_8|^2 \right) \, \varLambda_{2m,n+{1\over 2}} (R)- \left( O_8 \bar C_8 - C_8 \bar O_8 \right) \, \varLambda_{2m+1,n+{1\over 2}} (R)\,,
\cr}
$$
it is clear that bosons and fermions have {\it even} and {\it odd} momenta in the orbifold. 
It is however simple to relate the two settings \refs{\ADSi,\reviewi,\reviewii}: the conventional Scherk-Schwarz basis of Field Theory can be recovered letting $R^{\rm SS} \equiv \rho = {1\over 2} R$, so that
$$
\eqalign{
{\scr T} =& \left( |V_8 |^2 + |S_8 |^2 \right) \, \varLambda_{m ,2n} (\rho) - \left(S_8 \bar V_8 + V_8 \bar S_8 \right) \, \varLambda_{m+{1\over 2},2n} (\rho) 
\cr
&+ \left( |O_8|^2 + |C_8|^2 \right) \, \varLambda_{m,2n+1} (\rho)- \left( O_8 \bar C_8 - C_8 \bar O_8 \right) \, \varLambda_{m+{1\over 2},2n+1} (\rho)\,,
\cr}
$$
where bosons and fermions have indeed the correct quantum numbers.

Similar considerations apply also to the orientifolds, and in particular to the D-brane sector.
From eq. \openss\ one deduces that bosons and fermions have {\it even} and {\it odd} momenta along the brane world-volume. Therefore, to recover the standard Field Theory Kaluza-Klein spectrum, the effective length of the brane has to be halved $L_\|^{\rm SS} \equiv \lambda_\| = {1\over 2} L_\|$, similarly to the closed-string case. This has important consequences on the massless spectrum of open strings, induced by the unchanged quantisation condition \dirac . As a result of the halving of the fundamental cell, the wrapping numbers of the various stacks of branes change accordingly. In particular
\eqn\wrappings{
\eqalign{
(q_a \,,\, k_a ) &\to (\omega_a \,,\, \kappa_a ) =  (2 q_a \,,\, k_a )  \,,
\cr
(q_i \,,\, k_i ) &\to (\omega_i \,,\, \kappa_i ) = (q_i \,,\, {\textstyle{1\over 2}} k_i ) \,,
\cr}
}
and this identification makes it clear that the model in \torann\ and in \nonsusyann\ corresponds to a conventional Scherk-Schwarz (and M-theory) deformation of a configuration of branes with wrapping numbers $(\omega_\alpha \,,\, \kappa_\alpha)$ as in \wrappings .
Therefore, {\it non-chiral fermions from branes with even horizontal wrapping number $\omega_a$ stay massless, while those from branes with odd horizontal wrapping number $\omega_i$ get a tree-level mass proportional to $1/2\lambda_\|$}. Similar considerations can be straightforwardly generalised to the case of vertical or oblique shifts. This redefinition of the wrapping numbers also takes care of the additional factors of two and one-half in the multiplicities of the chiral sectors, so that the annulus 
$$
\eqalign{
{\scr A} =& \sum_{a=1}^{m_{\rm e}} \, N_a \bar N_a \, (V_8 - S_8) [{\textstyle{0\atop 0}}] \, 
\prod_{\varLambda =1}^3 \, P^\varLambda_m\, W^\varLambda_n
\cr
&+ \sum_{i=1}^{m_{\rm o}} \, N_i \bar N_i \, \left( V_8 [{\textstyle{0\atop 0}}]\, P^1_m -
S_8 [{\textstyle{0\atop 0}}] \, P^1_{m+{1\over 2}} \right) \, W^1_n \, 
\prod_{\varSigma =2,3} \, P^\varSigma_m \, W^\varSigma_n
\cr
&+ {\textstyle{1\over 2}} \, \sum_{\alpha =1}^{m_{\rm e} + m_{\rm o}} \left(
N_\alpha^2 \, (V_8 - S_8) [{\textstyle{\alpha\bar\alpha \atop 0}}] + \bar N_\alpha^2 \, (V_8 - S_8) [{\textstyle{\bar\alpha \alpha \atop 0}}] \right) \, {I_{\alpha\bar\alpha} \over \varUpsilon_1 [{\alpha\bar\alpha \atop 0}]}
\cr
&+ \sum_{{\alpha,\beta=1 \atop \beta<\alpha}}^{m_{\rm e} +m_{\rm o}} \left( N_\alpha \bar N_\beta (V_8 - S_8 ) 
[{\textstyle{\alpha \beta \atop 0}}] + \bar N_\alpha N_\beta (V_8 - S_8 ) [{\textstyle{\bar \alpha \bar\beta \atop 0}}]\right)\,
{I_{\alpha \beta} \over \varUpsilon_1 [{\textstyle{\alpha \beta\atop 0}}]}
\cr
&+ \sum_{{\alpha,\beta=1 \atop \beta<\alpha}}^{m_{\rm e}+m_{\rm o}} \left( N_\alpha  N_\beta (V_8 - S_8 ) 
[{\textstyle{\alpha\bar\beta\atop 0}}] + \bar N_\alpha \bar N_\beta (V_8 - S_8 ) 
[{\textstyle{\bar \alpha \beta\atop 0}}] \right) \,
{I_{\alpha\bar\beta} \over \varUpsilon_1 [{\textstyle{\alpha\bar\beta\atop 0}}]}
\cr}
$$
and M\"obius-strip
$$
{\scr M} = - {\textstyle{1\over 2}} \sum_{\alpha =1}^{m_{\rm e} + m_{\rm o}} \left(
N_\alpha \, (\hat V_8 - \hat S_8) [{\textstyle{\alpha\bar\alpha \atop 0}}] + \bar N_\alpha \, (\hat V_8 - \hat S_8) [{\textstyle{\bar\alpha \alpha \atop 0}}] \right) \, {K_\alpha \over \hat \varUpsilon_1 [{\alpha\bar\alpha \atop 0}]}
$$
amplitudes clearly define a freely acting deformation of the spectrum of conventional branes with wrapping numbers $(\omega_\alpha , \kappa_\alpha)$. In these equations for ${\scr A}$ and ${\scr M}$ the intersection numbers $I_{\alpha\beta}$ and $K_\alpha$ are clearly expressed in terms of $\omega_\alpha$ and $\kappa_\alpha$, and $m_{\rm e} \equiv n_{\rm o}$ ($m_{\rm o} \equiv n_{\rm e}$) counts the number of different stacks with $\omega_a$ even ($\omega_i$ odd). Moreover, additional non-chiral fermions arising from parallel branes in a particular $T^2$, can also be given a mass through a similar deformation. For instance, in the case of a pair of $\check \imath$ and $\check \jmath$ branes parallel in the first $T^2$ and with $\omega^1_{\check \imath} \equiv \omega^1_{\check \jmath}$ odd one finds the contribution
\eqn\nonchiral{
\left[ \left( N_{\check \imath} \bar N_{\check \jmath}  V_8 [{\textstyle{\check \imath \check\jmath \atop 0}}] + \bar N_{\check \imath}  N_{\check \jmath}  V_8 [{\textstyle{
\bar{\check \imath} \bar{\check\jmath} \atop 0}}]  \right) P^1_m - 
\left( N_{\check \imath} \bar N_{\check \jmath}  S_8 [{\textstyle{\check \imath \check\jmath \atop 0}}] + \bar N_{\check \imath}  N_{\check \jmath}  S_8 [{\textstyle{
\bar{\check \imath} \bar{\check\jmath} \atop 0}}]  \right) P^1_{m+{1\over 2}} \right] W^1_n \, {I^{\rm non\ chiral}_{\check\imath \check\jmath} \over \tilde \varUpsilon_1
[{\textstyle{\check \imath \check\jmath \atop 0}}]}
}
where as in the previous sub-section both $I^{\rm non\ chiral}_{\check\imath \check\jmath}$ and $\tilde\varUpsilon_1 [{\textstyle{\check \imath \check\jmath \atop 0}}]$ do not include terms from the first torus, and the argument of the internal SO(2) characters associated to it is automatically vanishing since $\phi^1_{\check\imath} - \phi^1_{\check\jmath} =0$. 

What happens if two pairs of branes are now parallel but in different $T^2$'s? One clearly has to deform the theory along both tori simultaneously. For definiteness, let us consider the case where  $\check \imath$ and $\check \jmath$ branes are parallel in the first $T^2$ and $\hat \imath$ and $\hat \jmath$ branes are parallel in the second $T^2$. Therefore, a deformation along the first torus of the type \nonchiral\ will give mass to the non-chiral fermions in the representation $(N_{\check \imath}\,,\, \bar N_{\check\jmath})$
proportional to $1/\lambda^1_{\|\,\check\imath}$, while a similar deformation along the second torus will give a tree-level mass to the non-chiral fermions in the representation
$(N_{\hat \imath}\,,\, \bar N_{\hat\jmath})$ proportional to $1/\lambda^2_{\|\,\hat\imath}$.
What about the gauginos in the adjoint representation? Here the situation is slightly more involved since we have now to split the set of branes in four categories depending on their horizontal wrapping numbers along the first and second $T^2$. If we label with $\alpha_1$ the branes with both $\omega^1$ and $\omega^2$ even, with $\alpha_2$ the branes with $\omega^1$ even and $\omega^2$ odd, with $\alpha_3$ the branes with $\omega_1$ odd and $\omega_2$ even, and finally with $\alpha_4$ those with both 
$\omega_1$ and $\omega_2$ odd the neutral dipole sector in the annulus amplitude reads
$$
\eqalign{
{\scr A}_{\rm dipole} =& \sum_{\alpha_1} N_{\alpha_1} \bar N_{\alpha_1} \, 
(V_8 - S_8) [{\textstyle{0\atop 0}}] \, \prod_{\varLambda=1}^3 P^\varLambda_m W^\varLambda_n 
\cr
&+ \sum_{\alpha_2} N_{\alpha_2} \bar N_{\alpha_2} \, 
\left( V_8 [{\textstyle{0\atop 0}}] \, P^2_m - S_8 [{\textstyle{0\atop 0}}] \, P^2_{m+{1\over 2}} \right)\, W^2_n \, \prod_{\varSigma=1,3} P^\varSigma_m W^\varSigma_n 
\cr
&+ \sum_{\alpha_3} N_{\alpha_3} \bar N_{\alpha_3} \, 
\left( V_8 [{\textstyle{0\atop 0}}] \, P^1_m - S_8 [{\textstyle{0\atop 0}}] \, P^1_{m+{1\over 2}} \right)\, W^1_n \, \prod_{\varSigma=1,2} P^\varSigma_m W^\varSigma_n 
\cr
&+\sum_{\alpha_3} N_{\alpha_3} \bar N_{\alpha_3} \, 
\left[ V_8 [{\textstyle{0\atop 0}}] \, \left( P^1_m \, P^2_m + P^1_{m+{1\over 2}} \, P^2_{m +{1\over 2}} \right) \right.
\cr
&\left. -
S_8 [{\textstyle{0\atop 0}}] \, \left( P^1_m \, P^2_{m+{1\over 2}} + P^1_{m+{1\over 2}} \, P^2_m \right) \right] W^1_n W^2_n \, P^3_m W^3_m
\cr}
$$
and clearly the $\alpha_1$ gauginos stay massless, the $\alpha_2$ gauginos get a mass proportional to $1/\lambda^2_{\|\, \alpha_2}$, the $\alpha_3$ gauginos get a mass proportional to $1/\lambda^1_{\|\, \alpha_3}$, while the $\alpha_4$ gauginos get a mass proportional to $\sqrt{1/(\lambda^1_{\|\, \alpha_4})^2+1/(\lambda^2_{\|\, \alpha_4})^2}$.
The generalisation to the case of deformations acting along the three $T^2$'s is then straightforward.

\subsec{An alternative Field Theory description}

To support and clarify of our results we can study a simple Field Theory problem where the Scherk-Schwarz deformation acts as usual along $y_1$ but now the various fields depend on the coordinate $y=\sqrt{(\omega \, y_1)^2 + (\kappa\, y_2)^2}$. 
To this end, we analyse the Kaluza-Klein spectrum of such fields subject to boundary conditions twisted by the operator $(-1)^F$. They correspond to excitations of neutral (parallel) strings, the only sector that admits zero modes and thus the only sector that can be affected by the deformation.  As in the previous sub-section, the pair of integers $(\omega, \kappa )$ on the $T^2$ define the line on which the fields live. 

On a rectangular $T^2$, the eigenfunctions of the two-dimensional Laplace and Dirac operators are simple plane waves
\eqn\eigenfunctions{
\varPhi _{p_1 , p_2 } (y_1 , y_2) \sim e^{2i\pi (p_1 y_1 + p_2 y_2 )}\,,
}
with the momenta determined by the periodicity conditions. As a result, a twist by $(-1)^F$ along the horizontal axis yields 
$$
p_1 = \left( m_1 + {\textstyle{1\over 2}} \, \varDelta_{\rm F} \right) {1\over R_1}\quad {\rm and} \qquad p_2 = {m_2 \over R_2} \,,
$$
with $\varDelta_{\rm F} = 0$ for space-time bosons, $\varDelta_{\rm F} = 1$ for space-time fermions, and $m_1$ and $m_2$ both integers. We can now determine the eigenfunctions of fields living on the straight line with 
\eqn\straightline{
\tan \phi = {\kappa R_2 \over \omega R_1}\,.
}
These can be obtained by \eigenfunctions\ by restricting the generic points $(y_1 , y_2)$ 
to lie on the straight line \straightline . The corresponding Kaluza-Klein spectrum 
$$
M^2 = \left( {\cos \phi \over \omega \, R_1}\right)^2 \left[ \left( m_1 + {\textstyle{1\over 2}} \, \varDelta_{\rm F} \right) \, \omega + m_2 \kappa \right]^2
$$
clearly shows that the twist $(-1)^F$ has a non-trivial effect on space-time fermions only if $\omega$ is an odd integer. 

Similar analysis can be performed in the case of twists along the vertical and diagonal directions, the only other choices compatible with the orientifold projection $\varOmega {\scr R}$. In these cases, the would-be gauginos in the dipole-string sector are lifted in mass only if $\kappa$ is odd for vertical twist, or both $\omega$ and $\kappa$ are odd for a diagonal twist.

\subsec{A deformed Standard-Model-like configuration}

As a simple application of Scherk-Schwarz deformations to models of some phenomenological relevance we can consider the intersecting-brane configuration of \ibanez. The Standard Model spectrum is there reproduced by four stacks of branes with gauge group
$$
G_{\rm CP} = {\rm U} (3)_a \times {\rm U} (2)_b \times {\rm U} (1)_c \times {\rm U} (1)_d \,.
$$
The additional Abelian factor is required in order to accommodate the right-handed leptons, that indeed emerge from open strings stretched between the ${\rm U}(1)_c$ and the ${\rm U} (1)_d$ branes. The four U(1)'s  are related to four global symmetries of the Standard Model: $Q_a$ is three times the barion number, $Q_d$ is minus the lepton number, $Q_c$ is twice the third component of the right-handed weak isospin, and finally $Q_b$ is a Peccei-Quinn symmetry. This lead the authors of \ibanez\ to identify the anomaly-free combination
$$
Q_{Y} = {\textstyle{1\over 6}} \, Q_a - {\textstyle{1\over 2}} \, Q_c + {\textstyle{1\over 2}} \, Q_d
$$
with the familiar hyper-charge. In order to reproduce the desired chiral spectrum the intersection numbers of the four stacks should be
$$
\eqalign{
I_{a b} &= 1 \,,
\cr 
I_{a c} &= -3 \,,
\cr
I_{b d} &= 0\,,
\cr
I_{c  d} &= -3 \,,
\cr}
\qquad \quad
\eqalign{
I_{a\bar b} &= 2 \,,
\cr 
I_{a \bar c} &= -3 \,,
\cr
I_{b \bar d} &= -3\,,
\cr
I_{c  \bar d} &= 3 \,.
\cr}
$$
It is not difficult to show that such intersection numbers can be obtained from branes with wrappings in table 2.

\vskip 10 pt
\vbox{\ninepoint
\settabs = 7\columns
\hskip 3.2truecm\vbox{\hrule width  10 truecm}
\+ 
&& $N_\alpha$ & $(\omega^1_\alpha \,,\, \kappa^1_\alpha)$ &
$(\omega^2_\alpha \,,\, \kappa^2_\alpha)$ &
$(\omega^3_\alpha \,,\, \kappa^3_\alpha)$ 
\cr
\hskip 3.2truecm\vbox{\hrule width  10 truecm}
\+
&& 3 & $({1\over \beta^1} , 0)$ & $(\omega^2_a , \epsilon \beta^2)$ & $({1\over \rho} , {1\over 2})$
\cr
\+
&& 2 & $(\omega^1_b , -\epsilon \beta^1 )$ & $({1\over\beta^2},0)$ & $(1, {3\over 2}\rho)$
\cr
\+
&& 1 & $(\omega^1_c , 3 \rho \epsilon \beta^1 )$ & $({1\over \beta^2} , 0)$ & $(0,1)$
\cr
\+
&& 1 & $({1\over \beta^1}, 0)$ & $(\omega_d^2 , - \beta^2 \epsilon \rho)$ & $(1, {3\over 2}\rho )$
\cr
\hskip 3.2truecm\vbox{\hrule width  10 truecm}
\cleartabs
\vskip 10pt
\centerline{\ninepoint{\bf Table 2.} D6-brane wrapping numbers giving rise to a SM-like spectrum, from \ibanez.}
\vskip 10pt
}
\noindent
With respect to ref. \ibanez, here we have followed our convention to use $\omega_\alpha$ and $\kappa_\alpha$ to denote the horizontal and vertical wrapping numbers, thus replacing the pairs $(n_\alpha \,,\, m_\alpha)$. Otherwise, $\epsilon =\pm 1$,  $\beta_i= 1 , {1\over 2}$ denotes (one minus) the real component of the complex structure that for $\Omega {\scr R} (-1)^{F_{\rm L}}$ orientifolds is discrete \discrete\ and takes the values zero or ${1\over 2}$, while $\rho=1,{1\over 3}$ is a parameter. The remaining $\omega$'s are not fully independent since their values enter in the assignment of the hyper-charge quantum numbers, and thus have to satisfy the constraint
$$
\omega^1_c = {\beta^2 \over 2\beta^1} \left( \omega_a^2 + 3\,\rho \, \omega_d^2 \right) \,,
$$
with $\omega^1_b$ undetermined. 

The choice 
$$
\rho = \beta^1 = \beta^2 = -\epsilon = 1\,, \quad \omega^2_a = 2\,, \quad \omega_b^1 = \omega_c^1 = 1 \,, \quad \omega_d^2 = 0\,,
$$
amounts to taking all the $\omega^1_\alpha$ to be odd (actually all equal to one), and thus
a Scherk-Schwarz deformation along the horizontal side of the first $T^2$ is enough to make all the would-be gaugino massive. Similarly, the non-chiral fermions in the $b\, d$ sector can be given a mass by deforming the third torus, where indeed the $b$ and $d$ branes are parallel. Moreover, non-chiral fermions in the $a\,\bar a$ and $d\,\bar d$ sectors are also massive since the branes are parallel in the first torus, while those in the $b\,\bar b$ and $c\,\bar c$ sectors, originating from branes parallel in the second $T^2$, can get a mass if the Scherk-Schwarz deformation acts also along the second torus. As a result, all massless non-chiral fermions can be properly lifted.

We should stress here that the wrapping numbers we have chosen, alike those suggested by the authors in \ibanez, do not satisfy the tadpole condition
$$
{3 \omega^2_a \over \rho\beta^1} + {2 \omega^1_b \over \beta^2} + {\omega^2_d \over \beta^1} = 16 \,.
$$
In principle, this failure can be overcome introducing extra hidden D6 branes with no intersection with the Standard-Model ones \ibanez. In this case, the above equation would be replaced by the more general one
$$
{3 \omega^2_a \over \rho\beta^1} + {2 \omega^1_b \over \beta^2} + {\omega^2_d \over \beta^1} + N_h \omega^1_h \omega^2_h \omega^3_h = 16 \,.
$$

\subsec{Deforming a three generations Pati-Salam model}

As a second example, let us examine the model presented in \fbralph. It is a four-dimensional, three generation, left-right symmetric standard model with gauge group
$$
G_{\rm CP} = {\rm U} (3)_a \times {\rm U} (2)_b \times {\rm U}(2)_c \times {\rm U} (1)_d
\,.
$$
It can be obtained engineering four stacks of D6 branes with intersection numbers as in table 3. To get the correct spectrum reported for completeness in table 4\footnote{$^\ast$}{See \fbralph, for more details on anomalous ${\rm U}(1)$'s, hyper-charge embedding, and some more phenomenological aspects.} only the second $T^2$ is chosen to be tilted. This also guarantees that all R-R tadpole conditions are now satisfied \fbralph.

In this example where all $\omega$'s are odd, regardless of the choice of (horizontal) Scherk-Schwarz coordinate, all adjoint fermions are lifted in mass. As for the remaining non-chiral fermions, they can be made massive deforming also the second torus.

\vskip 10pt

\vbox{\ninepoint
\settabs = 7\columns
\hskip 3.2truecm\vbox{\hrule width  10 truecm}
\+ 
&& $N_\alpha$ & $(\omega^1_\alpha \,,\, \kappa^1_\alpha)$ &
$(\omega^2_\alpha \,,\, \kappa^2_\alpha)$ &
$(\omega^3_\alpha \,,\, \kappa^3_\alpha)$ 
\cr
\hskip 3.2truecm\vbox{\hrule width  10 truecm}
\+
&& 3 & $(1,0)$ & $(1,0)$ & $(3,1)$
\cr
\+
&& 2 & $(1,1)$ & $(1,1)$ & $(1,0)$
\cr
\+
&& 2 & $(1,1)$ & $(1,-2)$ & $(1,0)$
\cr
\+
&& 1 & $(1,0)$ & $(1,-2)$ & $(3,1)$
\cr
\hskip 3.2truecm\vbox{\hrule width  10 truecm}
\cleartabs
\vskip 10pt
\centerline{\ninepoint{\bf Table 3.} D6-brane wrapping numbers for a left-right symmetric model, from \fbralph.}
\vskip 10pt
}

\vskip 10pt
\vbox{\ninepoint
\settabs=4\columns
\hskip 2.9truecm\vbox{\hrule width  11 truecm}
\+
&   ${\rm SU} (3) \times {\rm SU} (2)_{\rm L} \times {\rm SU} (2)_{\rm R} \times {\rm U} (1)^4 $ & & generations
\cr
\hskip 2.9truecm\vbox{\hrule width  11 truecm}
\+
&$\qquad\quad(3,2,1)_{(1,1,0,0)}$ & & $\qquad$ 2
\cr
\+
&$\qquad\quad(3,2,1)_{(1,-1,0,0)}$ & & $\qquad$ 1
\cr
\+
&$\qquad\quad(3^*,1,2)_{(-1,0,1,0)}$ & & $\qquad$ 2
\cr
\+
&$\qquad\quad(3^*,1,2)_{(-1,0,-1,0)}$ & & $\qquad$ 1
\cr
\+
&$\qquad\quad(1,2,1)_{(0,-1,0,1)}$ & & $\qquad$ 3
\cr
\+
&$\qquad\quad(1,1,2)_{(0,-1,0,-1)}$ & & $\qquad$ 3
\cr
\hskip 2.9truecm\vbox{\hrule width  11 truecm}
\cleartabs
\vskip 10pt
\centerline{\ninepoint{\bf Table 4.} Left-right symmetric chiral massless spectrum, from \fbralph.}
\vskip 10pt
}

\subsec{Scherk-Schwarz deformations on a tilted torus}

It is not difficult to generalise our previous results to the case of a tilted torus. 
As shown in \discrete, compatibility with the $\varOmega {\scr R}$ projection imposes constraints on the real part of the complex structure, whose quantum deformation does not belong any more to the  physical spectrum, but nevertheless can be given a non-trivial constant background value. This is obviously related by T duality to the more familiar case of the NS-NS $B$ field worked out in \toroidal\ for the simple world-sheet parity $\varOmega$. Describing magnetised or rotated branes in this tilted torus (see fig. 8) is quite straightforward, though some modifications are needed \refs{\augusto,\fbralph}. 
Firstly, if we define $(\omega , \kappa)$ as the number of times a brane wraps the two canonical cycles of the $T^2$, the quantisation condition on the angle reads
\eqn\newangle{
\tan\phi = {(\kappa + {1\over 2} \omega ) R_2 \over \omega R_1} = {\kappa R_2 \over \omega R_1} + {R_2\over 2\, R_1} \,,
}
with the additional contribution deriving from the fixed real part of the complex structure.
Moreover, the shear parameter enters both in the zero-mode sector, through a proper redefinition of the effective length of the brane
$$
L_\| = \sqrt{\left(\omega R_1\right)^2 + \left[\left(\kappa + {\textstyle{1\over 2}}\omega \right) R_2
\right]^2} \,,
$$
and in the frequencies of the string excitations that are shifted by the angle $\phi$ defined in \newangle . As usual, for any brane with angle $\phi$ and wrapping numbers $(\omega_\alpha , \kappa _\alpha)$ one has also to introduce image branes under the action of the orientifold operator $\varOmega{\scr R}$. They still have opposite angle $-\phi$, but now the wrapping numbers are $(\omega_\alpha , -\kappa_\alpha - \omega_\alpha )$, as a result of the tilting of the torus. Actually, all these changes are more easily taken into account after we label branes with wrapping numbers in a Cartesian basis.
In this respect, all the modifications can be summarised by noting that the wrapping numbers $(\omega_\alpha , \kappa_\alpha )$ for the case of a rectangular torus have to be replaced by  $(\omega_\alpha ,\kappa_\alpha + {1\over 2} \omega_\alpha)$ for the case of a tilted torus \fbralph.

\vbox{
\vskip 10pt
\epsfxsize 2truein
\centerline{\epsffile{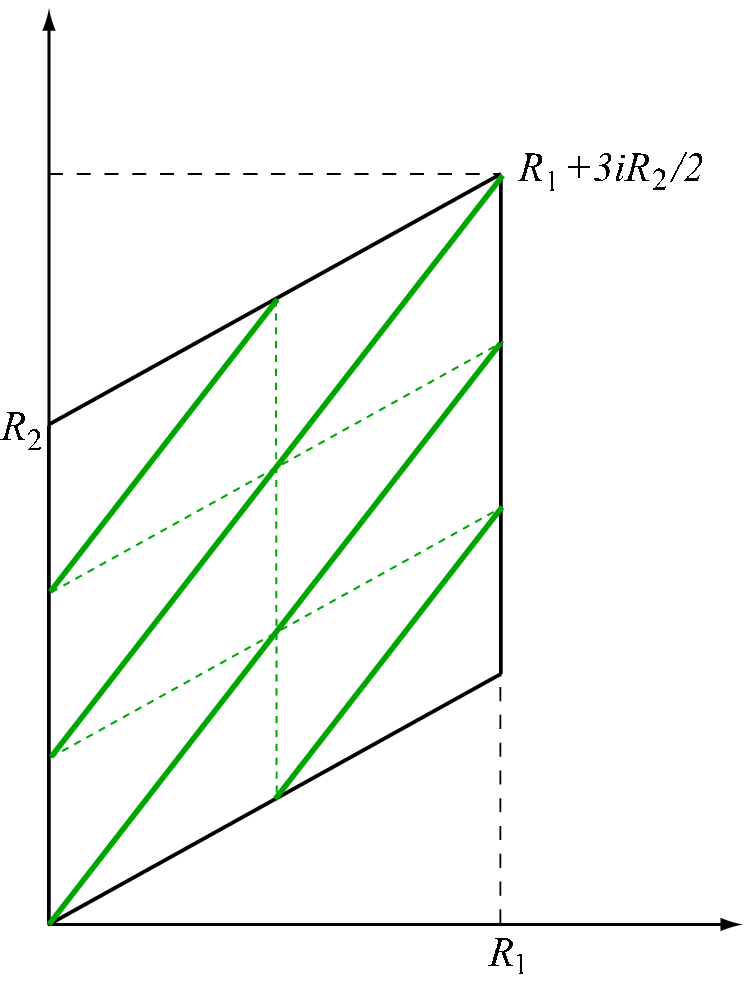}}
\vskip 10pt
\centerline{\ninepoint {\bf Figure 8.} A (3,2) brane wrapping a tilted torus.}
\vskip 10pt
}

With these modifications in mind, we can then straightforwardly generalise our previous results for a Scherk-Schwarz deformation acting along the horizontal axis to the case of a tilted torus: once more, it is the parity of $\omega_\alpha$ to discriminate between a trivial or non-trivial action on the neutral non-chiral fermions. 

\newsec{Six-dimensional orbifold models}

As a second application, we consider the non-supersymmetric type IIB string compactified on the $(T^2 \times T^2)/\bb{Z}_2$ with the Scherk-Schwarz deformation acting along the horizontal axis of the first $T^2$. 

\subsec{Preludio: the closed-string sector}

In the Scherk-Schwarz basis, the torus partition function reads \ADSi
\eqn\SStorus{
\eqalign{
{\scr T} =& {\textstyle{1\over 2}} \Biggl[ \left( |V_8|^2 + |S_8|^2 \right) \, \varLambda_{2n_1}^{(4,4)} 
- \left( S_8 \bar V_8 + V_8 \bar S_8 \right) \, \varLambda^{(4,4)}_{m_1 + {1\over 2},2 n_1}
+ \left( |O_8|^2 + |C_8|^2 \right) \, \varLambda^{(4,4)}_{2n_1 +1} 
\cr
& - 
\left( C_8 \bar O_8 + O_8 \bar C_8 \right) \, \varLambda^{(4,4)}_{m_1 +{1\over 2}, 2 n_1 + 1} 
+ \left( |V_4 O_4 - O_4 V_4 |^2 + |C_4 C_4 - S_4 S_4|^2 \right)\, \left| {2\eta\over\vartheta_2}\right|^4 \Biggr]
\cr
&+ {\textstyle{16 \over 4}} \, \left[ \left( |Q_s + Q_c|^2 + |Q_s ' + Q_c '|^2 \right) \, \left| {\eta\over\vartheta_4}\right|^4
+ \left( |Q_s - Q_c|^2 + |Q_s ' - Q_c '|^2 \right) \, \left| {\eta\over\vartheta_3}\right|^4
\right]\,,
\cr}
}
where $\varLambda^{(4,4)}$ is the four-dimensional Narain lattice with momenta $m_1 \,, \ldots \,, m_4$ and windings $n_1 \,, \ldots \,, n_4$. Moreover, $\varLambda_{2 n_1}^{(4,4)}$ indicates that the winding number $n_1$ is now an even integer, and similarly for the other terms. The $Q$'s are defined as usual by 
$$
\eqalign{
Q_o =& \, V_4 O_4 - C_4 C_4 \,,
\cr
Q_v =& \, O_4 V_4 - S_4 S_4 \,,
\cr
Q_s =& \, O_4 C_4 - S_4 O_4 \,,
\cr
Q_c =& \, V_4 S_4 - C_4 V_4 \,,
\cr}
\qquad
\eqalign{
Q_o ' =& \, V_4 O_4 - S_4 S_4 \,,
\cr
Q_v ' =& \, O_4 V_4 - C_4 C_4 \,,
\cr
Q_s '=& \, O_4 S_4 - C_4 O_4 \,,
\cr
Q_c '=& \, V_4 C_4 - S_4 V_4 \,,
\cr}
$$
where we have introduced additional $Q$'s that will play a role in the following. 
The Klein bottle amplitude obtained by the $\varOmega {\scr R} (-1)^{F_{\rm L}}$ projection reads \ADSi
$$
\eqalign{
{\scr K} =& {\textstyle{1\over 4}} (V_8 - S_8 ) \left( P_{m}^1 W_{n}^2 P_{m}^3 W_{n}^4 + W_{2 n}^1 P_{m}^2 W_{n}^3 P_{m}^4 \right)
+ {\textstyle{1\over 4}} (O_8 - C_8 ) \, W_{2n+1}^1 P_{m}^2 W_{n}^3 P_{m}^4
\cr
& + 4 \left( Q_s + Q_c + Q_s ' + Q_c '\right) \, \left( {\eta \over \vartheta_4 }\right)^2 \,.
\cr}
$$
After an $S$-modular transformation to the transverse channel, and adapting to our six-dimensional example the definition in \radii , the massless contributions to the NS-NS and R-R tadpoles
$$
\Kt_0 \sim  8 \, V_4 O_4 \, \left( {\bf R} + {1\over {\bf R}}\right)^2
+ 8\, O_4 V_4 \, \left( {\bf R} - {1\over {\bf R}}\right)^2
- 8 \, (C_4 C_4 + S_4 S_4) \, {\bf R}^2
$$
clearly suggest that the O7-planes stretched along the vertical axis of the two $T^2$'s come in conjugate pairs and thus do not induce any R-R charge for the corresponding eight-form potential. The configuration of O7-planes stretched along the horizontal axis has instead a non-trivial charge that has to be compensated by suitable brane configurations \ADSi.
Since we are altering the nature of some orientifold planes, we cannot expect that the 
new open-string spectrum might be obtained from the undeformed one by giving masses to the fermions in the adjoint of the ${\rm U} (N_i)$ gauge group factors. In fact, this configuration would not solve any longer the untwisted R-R tadpole conditions, and thus new vacuum configurations have to be found. 

\vbox{
\vskip 10pt
%\centerline{\includegraphics[width=2in]{zero-mode.pdf}}
\epsfxsize 10truecm
\centerline{\epsffile{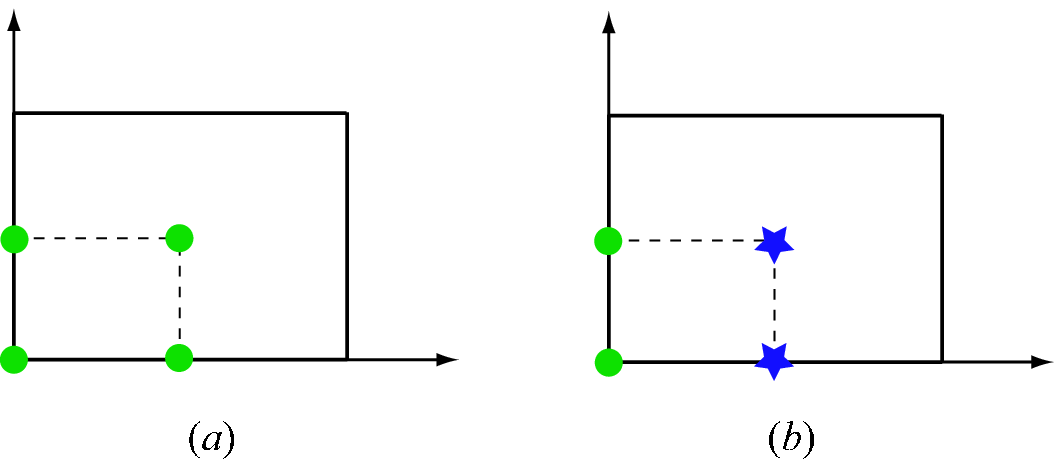}}
\vskip 10pt
\centerline{\ninepoint {\bf Figure 9.} Fixed points for a $\bb{Z}_2$ orbifold without $(a)$ and with $(b)$ Scherk-Schwarz deformation.}
\vskip 10pt
}

Before we proceed with the construction of the open-string sector, let us pause for a moment and try to extract some useful information about the geometry of this Scherk-Schwarz-deformed $T^4/\bb{Z}_2$. To this end it is convenient to stare at the twisted sector in \SStorus. The appearance of new characters $Q_{s,c}'$ and the modified multiplicity clearly spells out that the original sixteen fixed points of the geometrical $\bb{Z}_2$ orbifold model are now split into two sets of eight, with different GSO projections in the corresponding twisted sector. Moreover, one can also read from the terms $Q_s - Q_c$ and $Q_s' - Q_c'$ that the fixed points act differently on the internal quantum numbers. Indeed, if we call $\xi_\mu$ and $x_\mu$ ($\mu =1,\ldots , 8$) the fixed points in the two sets, the eight $\xi_\mu$ project the spectrum with respect the geometrical $\bb{Z}_2$ while the eight $x_\mu$ involve also the action of the Scherk-Schwarz deformation $(-1)^F$, for an overall $g\, (-1)^F$ projection, $g$ being the $\bb{Z}_2$ generator. This is clear both from the kind of boundary conditions we are imposing and from the structure of the modular-invariant partition function, where the projection in $Q_s' - Q_c'$ involves and additional $(-1)^F$ with respect to that in $Q_s - Q_c$. We can actually say more about the geometrical distribution of these fixed points. In our factorised $T^4 = T^2 \times T^2$ example, in fact, we have chosen to impose Scherk-Schwarz boundary conditions along the horizontal axis of the first torus. As a result, the structure of the fixed points in the second torus cannot be affected. In the first torus, instead, we have the configuration depicted in figure 9, since, as previously reminded, we are deforming the theory by changing the boundary conditions along the horizontal axis. Therefore, we can conclude that our $T^4$ is of the form $T^2_{(b)} \times T^2_{(a)}$, where $T^2_{(a)}$ and $T^2_{(b)}$ denote the undeformed and the deformed torus as in figure 9. The two sets of fixed points are then 
\eqn\fixedpoints{
\eqalign{
\vec\xi =& \left\{ (0,0,0,0) , (0,0,{\textstyle{1\over 2}} ,0) , (0,0,0, {\textstyle{1\over 2}}) , 
(0,0,{\textstyle{1\over 2}},{\textstyle{1\over 2}}) , \right.
\cr
&\left. (0,{\textstyle{1\over 2}},0,0), (0,{\textstyle{1\over 2}},{\textstyle{1\over 2}} ,0) , (0,{\textstyle{1\over 2}},0, {\textstyle{1\over 2}}) , 
(0,{\textstyle{1\over 2}},{\textstyle{1\over 2}},{\textstyle{1\over 2}}) \right\} \,,
\cr
\vec x =& \left\{ ({\textstyle{1\over 2}},0,0,0), ({\textstyle{1\over 2}},0,{\textstyle{1\over 2}} ,0) , ({\textstyle{1\over 2}},0,0, {\textstyle{1\over 2}}) , 
({\textstyle{1\over 2}},0,{\textstyle{1\over 2}},{\textstyle{1\over 2}}), \right.
\cr
&\left. ({\textstyle{1\over 2}},{\textstyle{1\over 2}},0,0), ({\textstyle{1\over 2}},{\textstyle{1\over 2}},{\textstyle{1\over 2}} ,0) , ({\textstyle{1\over 2}},{\textstyle{1\over 2}},0, {\textstyle{1\over 2}}) , 
({\textstyle{1\over 2}},{\textstyle{1\over 2}},{\textstyle{1\over 2}},{\textstyle{1\over 2}}) \right\} \,.
\cr}
}

\subsec{Intermezzo: the geometry of orthogonal branes}

We can now proceed to include the D-branes, and to better appreciate the construction of the open-string sector we first review the model presented in \ADSi, with the obvious replacement of D9 and D5 with two sets of orthogonal D7 branes. To reproduce their geometry we distribute the branes as in figure 10, that, after two T-dualities along the vertical axis of the two $T^2$'s correspond indeed to space-filling D9 branes together with a set of D5 branes sitting at the fixed point $\xi_1 = (0,0,0,0)$ and a set of D5 anti-branes sitting at the fixed point $x_1 = ({1\over 2},0,0,0)$. 

\vbox{
\vskip 10pt
%\centerline{\includegraphics[width=2in]{zero-mode.pdf}}
\epsfxsize 10truecm
\centerline{\epsffile{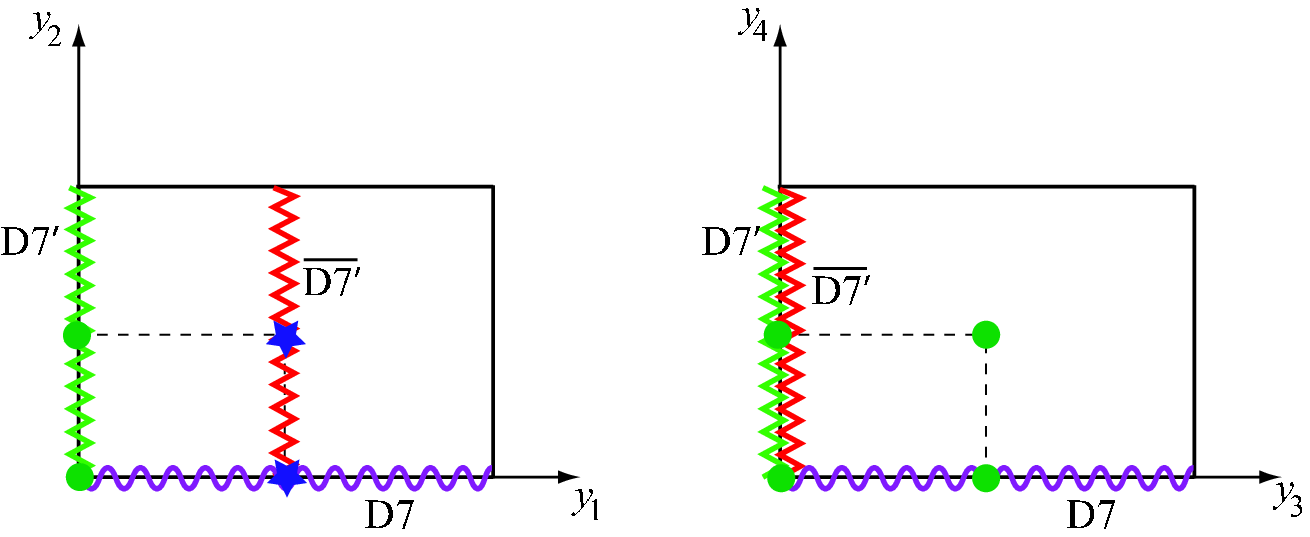}}
\vskip 10pt
\noindent{\ninepoint {\bf Figure 10.} The geometry of D-branes. A wavy line represents D7 branes while zig-zag lines represents ${\rm D}7'$ and $\overline{{\rm D}7'}$.}
\vskip 10pt
}

The direct-channel annulus in \ADSi\ is then
$$
\eqalign{
{\scr A} =& {\textstyle{1\over 4}} \Biggl\{ N_7^2 \left( V_8 \, P^1_m - S_8 \, P^1_{m+{1\over 2}} \right) W^2_n P^3_m W^4_n + \left( N_{7'}^2 + N_{\overline{7'}}^2 \right) (Q_o + Q_v )\, W_n^1 P_m^2 W_n^3 P_m^4
\cr
&+ 2 N_{7'} N_{\overline{7'}}\, (O_8 - C_8) W^1_{n+{1\over 2}} P^2_m W^3_n P^4_m +
R^2_7\, (V_4 O_4 - O_4 V_4 ) \left( {2\eta\over\vartheta_2}\right)^2
\cr
&+ \left[ R_{7'}^2 (Q_o - Q_v) + R_{\overline{7'}}^2 (Q_o ' - Q_v ') \right] \left( {2\eta \over \vartheta_2}\right)^2
\cr
&+ 2 \left[ N_7 N_{7'}\, (Q_s + Q_c) + N_7 N_{\overline{7'}} (Q_s ' + Q_c ' ) \right] \left( {\eta\over \vartheta_4} \right)^2
\cr
&+ 2 \left[ R_7 R_{7'} (Q_s - Q_c ) + R_7 R_{\overline{7'}} (Q_s ' - Q_c ' ) \right] \left( {\eta \over \vartheta_3}\right)^2 \Biggr\}\,.
\cr}
$$
Any single term in this expression has a clear rational. The terms in the first line simply correspond to open strings living on a given D-brane, with the important difference that D7 branes stretching along the horizontal $y_1$ axis now have a deformed non-supersymmetric spectrum. The second line is also quite standard: the first term encodes the spectrum of a brane-anti-brane system with the two ${\rm D}7'$ and $\overline{D 7'}$ branes separated along $y_1$, while the second contribution pertains to the orbifold projection on the horizontal D7 branes. Also the fourth line is quite standard: it corresponds to open strings with mixed Neumann-Dirichlet boundary conditions along the four-compact directions. The second term has different GSO projections since the strings stretch between a brane and an anti-brane. More subtle are the third and fifth lines. They both enforce the orbifold projection on open-strings, but in a different way. In fact, while the first term in the third line clearly corresponds to the conventional geometrical $\bb{Z}_2$ projection, the second term involves and additional minus sign in the Ramond sector, and indeed corresponds to the action of the element $g (-1)^F$. This is consistent with, and in fact dictated by, the fact that ${\rm D}7'$ branes sit on top of conventional $g$-fixed points, while the $\overline{{\rm D}7'}$ branes pass through points fixed under the action of $g (-1)^F$. Clearly, similar considerations also apply to the terms in the fifth line that describe the symmetrisation of $7\, 7'$ and $7\,\overline{7'}$ states that indeed live at the fixed points $\xi_1$ and $x_1$.
As usual, the different orbifold projections in the direct channel translate into exchanges of closed-string twisted states in the transverse channel, whose massless tadpoles clearly spell out the distribution of branes among the fixed points
$$
\At_{\rm tw} \sim 2^{-5}\, \left\{ \left[ \left(R_7 - R_{7'} \right)^2 + R^2_7 + 3\, R^2_{7'} \right]\, Q_s
+  \left[ \left(R_7 - R_{\overline{7'}} \right)^2 + R^2_7 + 3\, R^2_{\overline{7'}} \right]\, Q_s '
\right\} \,,
$$
compatible with the geometry in figure 10.

\vbox{
\vskip 10pt
%\centerline{\includegraphics[width=2in]{zero-mode.pdf}}
\epsfxsize 10truecm
\centerline{\epsffile{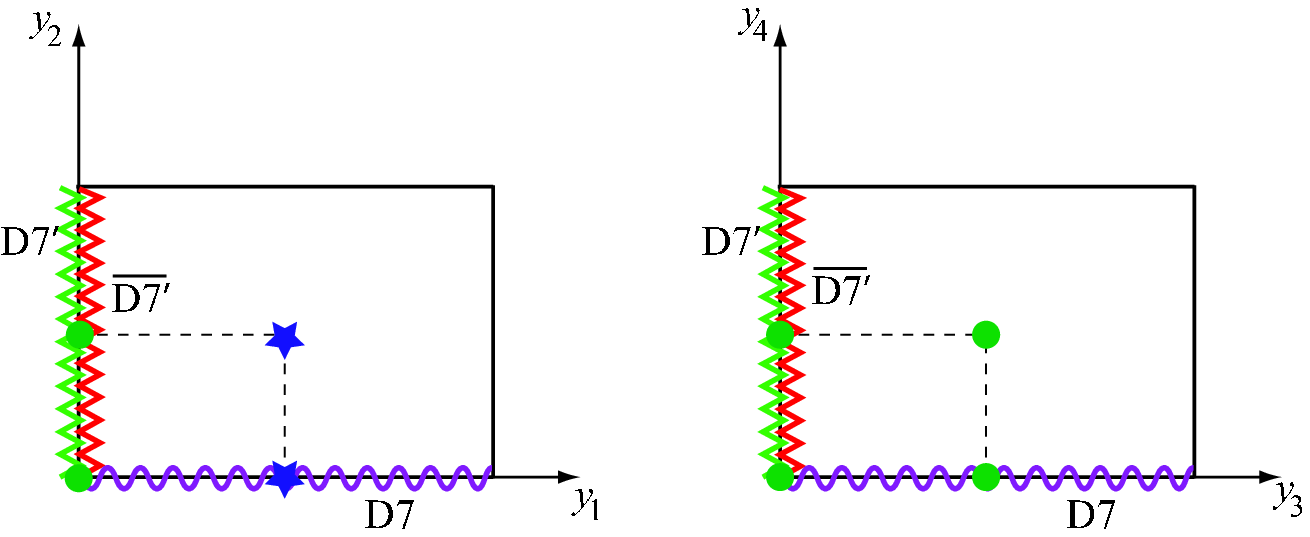}}
\vskip 10pt
\noindent{\ninepoint {\bf Figure 11.} The geometry of D-branes. A wavy line represents D7 branes while zig-zag lines represents ${\rm D}7'$ and $\overline{{\rm D}7'}$.}
\vskip 10pt
}

What about if we distribute branes differently? Let us consider for instance the configuration in figure 11. Now all the branes pass through the conventional $g$-fixed points, and thus we do not expect in the amplitude terms like $Q_o' - Q_v '$. Indeed, the new annulus amplitude
$$
\eqalign{
{\scr A} =& {\textstyle{1\over 4}} \Biggl\{ N_7^2 \, \left( V_8 \, P^1_m - S_8 \, P^1_{m+{1\over 2}} \right) W^2_n P^3_m W^4_n + \left( N_{7'}^2 + N_{\overline{7'}}^2 \right) (Q_o + Q_v) \, W^1_n P^2_m W^3_n P^4_m 
\cr
&+ 2 N_{7'} N_{\overline{7'}} \, (O_8 - C_8) \, W^1_n P^2_m W^3_n P^4_m + R_7^2\, (V_4 O_4 - O_4 V_4 ) \left( {2\eta\over \vartheta_2}\right)^2
\cr
&+ \left[ \left( R_{7'}^2 + R_{\overline{7'}}^2 \right) (Q_o - Q_v)  + 2 R_{7'} R_{\overline{7'}} \,
(O_4 O_4 - V_4 V_4 - S_4 C_4 + C_4 S_4)\right] \left( {2\eta\over \vartheta_2}\right)^2
\cr
&+ 2 \left[ N_7 N_{7'}\, (Q_s + Q_c ) + N_7 N_{\overline{7'}}\, (Q_s '+ Q_c ')  \right] \left( {\eta\over\vartheta_4} \right)^2
\cr
&+ 2 \left[ R_7 R_{7'}\, (Q_s - Q_c ) + R_7 R_{\overline{7'}}\, (- O_4 S_4 + V_4 C_4 - C_4 O_4 + S_4 V_4 ) \right]  \left( {\eta\over\vartheta_3} \right)^2
\cr}
$$
only involves the geometrical $\bb{Z}_2$ projector, and yields twisted tadpoles in the transverse channel
$$
\eqalign{
\At _{\rm tw} \sim& 2^{-5} \left\{ 2 R^2_7\, Q_s ' + \left[ \left( R_7 - R_{7'} + R_{\overline{7'}} \right)^2 + R^2_7 + 3 (R_{7'} - R_{\overline{7'}} )^2 \right] \, O_4 C_4 \right.
\cr
&- \left. \left[ \left( R_7 - R_{7'} - R_{\overline{7'}} \right)^2 + R^2_7 + 3 (R_{7'} + R_{\overline{7'}} )^2 \right] \, S_4 O_4 \right\}
\cr}
$$
that clearly display the distribution in figure 11. 

Similarly, one could opt for a different distribution of branes along the compact $T^4$, that would consequently imply different projections according to the nature of the fixed points crossed by the branes. This would then reflect in twisted tadpoles compatible with the chosen geometry.

\subsec{Crescendo: rotating the branes}

We can now generalise the previous construction to the case of rotated branes, where some care is needed to identify the exact location of brane intersections. In the simple case where all branes cross the origin of the two tori, it was shown in \fbralph\ that any brane always passes through four fixed points of the $T^4/\bb{Z}_2$. Moreover, we can easily identify 
the four points from the wrapping numbers $(\omega_\alpha^\varLambda , \kappa_\alpha^\varLambda)$, since for a single $T^2$ branes fall into three different equivalence classes: $\omega_\alpha$ even and $\kappa_\alpha$ odd, $\omega_\alpha$ odd and $\kappa_\alpha$ even, $\omega_\alpha$ and $\kappa_\alpha$ both odd and co-prime. Focussing our attention to the first $T^2$, the only one that actually matters to identify the correct projection, one has then the cases reported in figure 12. 

\vbox{
\vskip 10pt
%\centerline{\includegraphics[width=2in]{zero-mode.pdf}}
\epsfxsize 10truecm
\centerline{\epsffile{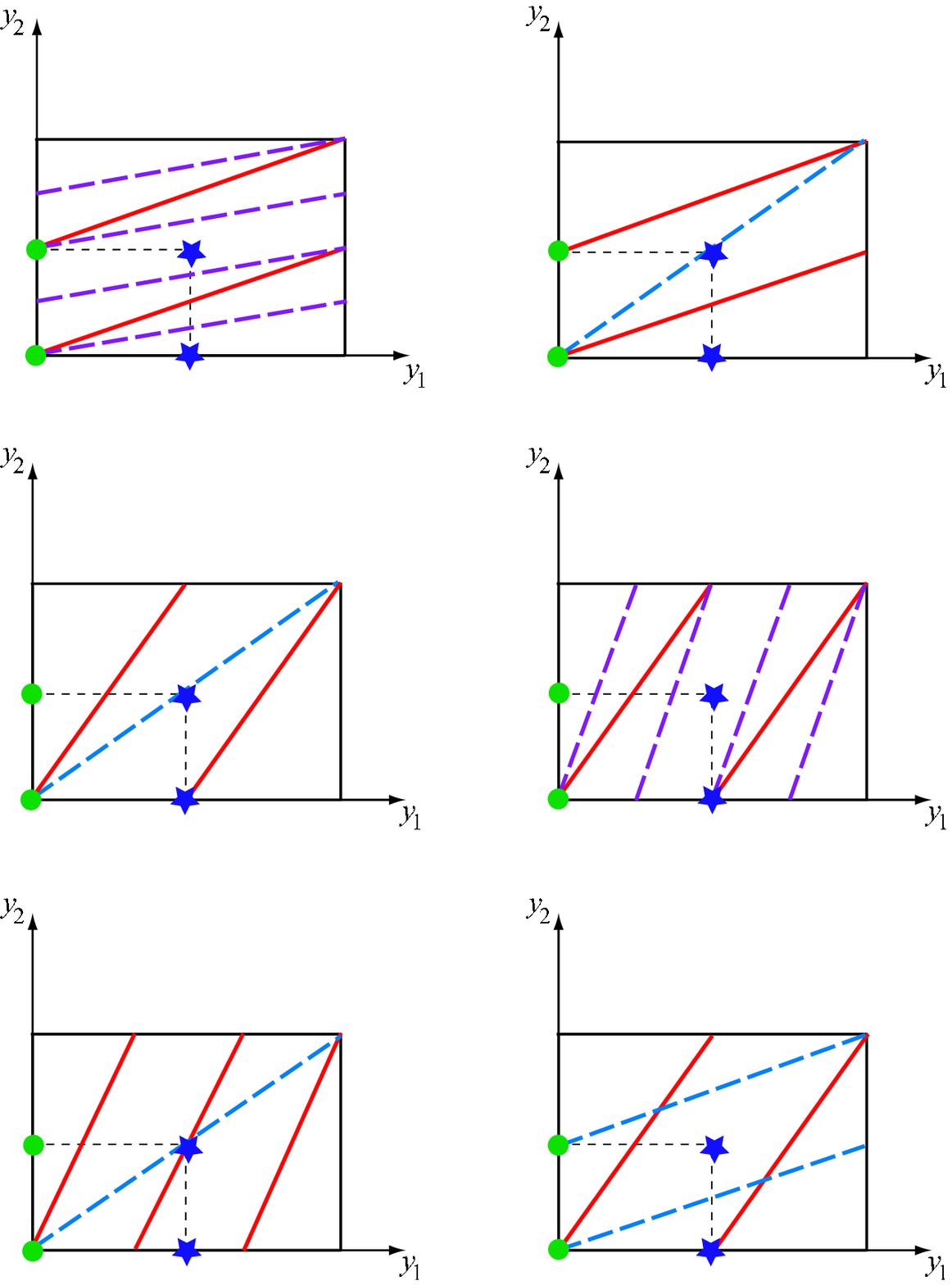}}
\vskip 10pt
\centerline{\ninepoint {\bf Figure 12.} Branes with various wrapping numbers on a $(b)$ torus.}
\vskip 10pt
}

The annulus amplitude associated to the closed-string sector in sub-section 5.1 is then
$$
\eqalign{
{\scr A}_{\alpha \alpha} =&  {\textstyle{1\over 2}}
\sum_{a=1}^{m_{\rm e}} N_a \bar N_a \, (Q_o + Q_v) [{\textstyle{0\atop 0}}] \, P_m^1 W_n^1 P_m^2 W_n^2 + {\textstyle{1\over 2}} \sum_{a=1}^{m_{\rm e}} N_a \bar N_a \, (Q_o - Q_v) [{\textstyle{0\atop 0}}] \, {4\over \varUpsilon_2 [{\textstyle{0\atop 0}}]} 
\cr
&+ {\textstyle{1\over 2}} \sum_{i=1}^{m_{\rm o}} N_i \bar N_i \, \left( V_8  [{\textstyle{0\atop 0}}] \, P_{m}^1 - S_8 [{\textstyle{0\atop 0}}]  \, P_{m+{1\over 2}}^1 \right)  W_n^1 P_m^2 W_n^2
\cr
&+ {\textstyle{1\over 2}} \sum_{i=1}^{m_{\rm o}} N_i \bar N_i \, ( V_4 O_4 - O_4 V_4 )
[{\textstyle{0\atop 0}}] \, {4\over \varUpsilon_2 [{\textstyle{0\atop 0}}]} \,,
\cr}
$$
$$
\eqalign{
{\scr A}_{\alpha \bar\alpha} +{\scr A}_{\bar \alpha \alpha} =& {\textstyle{1\over 4}} \sum_{\alpha =1}^{m_{\rm o}+m_{\rm e}} \left( N_\alpha^2 (Q_o + Q_v) [{\textstyle{\alpha\bar\alpha\atop 0}}] + \bar N_\alpha^2 (Q_o + Q_v) [{\textstyle{\bar \alpha \alpha \atop 0}}] \right) \, 
{I_{\alpha\bar\alpha}  \over \varUpsilon_1 [{\textstyle{\alpha\bar\alpha\atop 0}}]}
\cr
&- {\textstyle{1\over 4}} \sum_{a=1}^{m_{\rm e}} \left( N_a^2 (Q_o - Q_v) [{\textstyle{a\bar a\atop 0}}] + \bar N_a^2 (Q_o - Q_v) [{\textstyle{\bar a a  \atop 0}}] \right) \, 
{4 \over \varUpsilon_2 [{\textstyle{a\bar a\atop 0}}]} 
\cr
&- {\textstyle{1\over 4}} \sum_{i=1}^{m_{\rm o}} \left( N_i^2 (Q_o - Q_v) [{\textstyle{i\bar \imath\atop 0}}] + \bar N_i^2 (Q_o - Q_v) [{\textstyle{\bar \imath i  \atop 0}}] \right) \, 
{2 \over \varUpsilon_2 [{\textstyle{i\bar \imath\atop 0}}]} 
\cr
&- {\textstyle{1\over 4}} \sum_{i=1}^{m_{\rm o}} \left( N_i^2 (Q_o '- Q_v ') [{\textstyle{i\bar \imath\atop 0}}] + \bar N_i^2 (Q_o '- Q_v ') [{\textstyle{\bar \imath i  \atop 0}}] \right) \, 
{2 \over \varUpsilon_2 [{\textstyle{i\bar \imath\atop 0}}]} \,,
\cr}
$$
$$
\eqalign{
{\scr A}_{\alpha \beta } + {\scr A}_{\bar\alpha \bar \beta} =& {\textstyle{1\over 2}}
\sum_{{\alpha,\beta=1 \atop \beta<\alpha}}^{m_{\rm o}+m_{\rm e}} \left( N_\alpha \bar N_\beta (Q_o + Q_v ) 
[{\textstyle{\alpha \beta\atop 0}}] + \bar N_\alpha N_\beta (Q_o + Q_v ) [{\textstyle{\bar \alpha \bar\beta\atop 0}}]\right)\,
{I_{\alpha \beta} \over \varUpsilon_1 [{\textstyle{\alpha \beta\atop 0}}]}
\cr
&+ {\textstyle{1\over 2}}
\sum_{{a , b=1 \atop b<a}}^{m_{\rm e}} \left( N_a \bar N_b (Q_o - Q_v ) 
[{\textstyle{ab\atop 0}}] + \bar N_a N_b (Q_o - Q_v ) [{\textstyle{\bar a \bar b\atop 0}}]\right)\,
{2 I_{ab} ' \over \varUpsilon_2 [{\textstyle{ab\atop 0}}]} 
\cr
&+ {\textstyle{1\over 2}}
\sum_{a =1}^{m_{\rm e}} \sum_{i =1}^{m_{\rm o}}  \left( N_a \bar N_i (Q_o - Q_v ) 
[{\textstyle{ai\atop 0}}] + \bar N_a N_i (Q_o - Q_v ) [{\textstyle{\bar a \bar \imath \atop 0}}]\right)\,
{I_{ai} ' \over \varUpsilon_2 [{\textstyle{ai\atop 0}}]} 
\cr
&+ {\textstyle{1\over 2}}
\sum_{{i , j=1 \atop j<i}}^{m_{\rm o}} \left( N_i \bar N_j (Q_o - Q_v ) 
[{\textstyle{ij\atop 0}}] + \bar N_i N_j (Q_o - Q_v ) [{\textstyle{\bar \imath \bar \jmath\atop 0}}]\right)\,
{I_{ij} ' \over \varUpsilon_2 [{\textstyle{ij\atop 0}}]} 
\cr
&+ {\textstyle{1\over 2}}
\sum_{{i , j=1 \atop j<i}}^{m_{\rm o}} \left( N_i \bar N_j (Q_o ' - Q_v ') 
[{\textstyle{ij\atop 0}}] + \bar N_i N_j (Q_o '- Q_v ') [{\textstyle{\bar \imath \bar \jmath\atop 0}}]\right)\,
{I_{ij} '' \over \varUpsilon_2 [{\textstyle{ij\atop 0}}]} 
\,,
\cr}
$$
and, finally, 
$$
\eqalign{
{\scr A}_{\alpha\bar\beta} + {\scr A}_{\bar \alpha \beta} =& {\textstyle{1\over 2}} 
\sum_{{\alpha,\beta=1 \atop \beta<\alpha}}^{m_{\rm o}+m_{\rm e}} \left( N_\alpha  N_\beta (Q_o + Q_v ) [{\textstyle{\alpha\bar\beta\atop 0}}] + \bar N_\alpha \bar N_\beta (Q_o + Q_v ) 
[{\textstyle{\bar \alpha \beta\atop 0}}] \right) \,
{I_{\alpha\bar\beta} \over \varUpsilon_1 [{\textstyle{\alpha\bar\beta\atop 0}}]}
\cr
&- {\textstyle{1\over 2}}
\sum_{{a , b=1 \atop b<a}}^{m_{\rm e}} \left( N_a N_b (Q_o - Q_v ) 
[{\textstyle{a\bar b\atop 0}}] + \bar N_a \bar N_b (Q_o - Q_v ) [{\textstyle{\bar a b\atop 0}}]\right)\,
{2 I_{a\bar b} ' \over \varUpsilon_2 [{\textstyle{a\bar b\atop 0}}]} 
\cr
&- {\textstyle{1\over 2}}
\sum_{a =1}^{m_{\rm e}} \sum_{i =1}^{m_{\rm o}}  \left( N_a  N_i (Q_o - Q_v ) 
[{\textstyle{a\bar \imath\atop 0}}] + \bar N_a \bar N_i (Q_o - Q_v ) [{\textstyle{\bar a i \atop 0}}]\right)\,
{I_{a\bar \imath} ' \over \varUpsilon_2 [{\textstyle{a\bar \imath\atop 0}}]} 
\cr
&- {\textstyle{1\over 2}}
\sum_{{i , j=1 \atop j<i}}^{m_{\rm o}} \left( N_i N_j (Q_o - Q_v ) 
[{\textstyle{i\bar \jmath\atop 0}}] + \bar N_i \bar N_j (Q_o - Q_v ) [{\textstyle{\bar \imath j \atop 0}}]\right)\,
{I_{i\bar \jmath} ' \over \varUpsilon_2 [{\textstyle{i\bar \jmath\atop 0}}]} 
\cr
&- {\textstyle{1\over 2}}
\sum_{{i , j=1 \atop j<i}}^{m_{\rm o}} \left( N_i N_j (Q_o '- Q_v ') 
[{\textstyle{i\bar \jmath\atop 0}}] + \bar N_i \bar N_j (Q_o '- Q_v ') [{\textstyle{\bar \imath j \atop 0}}]\right)\,
{I_{i\bar \jmath} '' \over \varUpsilon_2 [{\textstyle{i\bar \jmath\atop 0}}]} 
\,.
\cr}
$$
Here we have used the obvious notation for the $Q [{\alpha\beta\atop\gamma\delta}]$ where the internal SO(4) characters are decomposed in terms of ${\rm SO}(2) \times {\rm SO} (2)$ ones, and
$$
\varUpsilon_2 [{\textstyle{\alpha\beta \atop \gamma\delta}}] = \prod_{\varLambda =1,2} \, {\vartheta_2 (\zeta^\varLambda |\tau ) \over \eta (\tau) } \, e^{2i \pi \zeta^\varLambda} 
\,,
$$
accounts for the $\bb{Z}_2$ orbifold generator acting on the world-sheet bosons.
Finally, $I_{\alpha\beta}$ denotes as usual the number of intersections of branes with wrapping numbers $(\omega_\alpha^\varLambda \,,\, \kappa_\alpha^\varLambda)$ and $(\omega_\beta^\varLambda \,,\, \kappa_\beta^\varLambda)$, while
$$
I'_{\alpha\beta} =  1 + \varPi (\omega_\alpha^2 - \omega_\beta^2 ) \,\varPi 
(\kappa_\alpha^2 - \kappa_\beta^2 ) 
$$
and 
$$
I''_{\alpha\beta} = \varPi (\omega_\alpha^1 +1 ) \varPi ( \omega_\beta^1 +1) \,\varPi 
(\kappa_\alpha^1 - \kappa_\beta^1 ) \, I'_{\alpha\beta} 
$$
count the number of intersections that actually coincide with the $x_p$ and $\xi_p$ fixed points in \fixedpoints , with
$$
\varPi ( \mu ) = {\textstyle{1\over 2}} \sum_{\epsilon=0,1} e^{i\pi\epsilon \mu}\,, \qquad \mu\in\bb{Z}\,,
$$
a $\bb{Z}_2$ projector. 

Finally, the M\"obius-strip amplitude
\eqn\moeborb{
\eqalign{
{\scr M} =& - {\textstyle{1\over 4}} \sum_{\alpha =1}^{m_{\rm o} + m_{\rm e}} \, 
\left[ \left(
N_\alpha \, (\hat Q _o + \hat Q_v ) [{\textstyle{\alpha \bar\alpha \atop 0}}] + \bar N_\alpha
\, (\hat Q _o + \hat Q_v ) [{\textstyle{\bar \alpha \alpha \atop 0}}] \right) \,
{K_\alpha \over \hat \varUpsilon_1  [{\textstyle{\alpha \bar\alpha \atop 0}}]} \right.
\cr 
&\left. - \left( N_\alpha \, (\hat V_4 \hat O_4 - \hat O_4 \hat V_4 ) [{\textstyle{\alpha \bar\alpha \atop 0}}] + \bar N_\alpha \, (\hat V_4 \hat O_4 - \hat O_4 \hat V_4) [{\textstyle{\bar \alpha\alpha \atop 0}}] \right) \,{ J_\alpha \over \hat \varUpsilon_2  [{\textstyle{\alpha \bar\alpha \atop 0}}]} \right]
\cr}
}
takes now into account also the (anti-)symmetrisation of states that live at intersections sitting on the vertical O7 planes, whose number is given by
$$
J_\alpha = \prod_{\varLambda=1,2} \, 2\, \omega_\alpha^\varLambda \,.
$$
Notice that the second line in ${\scr M}$ does not involve fermions. This has a simple explanation in terms of the relative R-R charges of vertical O-planes and branes whose intersections lie on them. We already commented that the absence in $\Kt_0$ of R-R tadpoles proportional to ${\bf R}^{-1}$ clearly spells out that the O7 planes passing through the points $(0,0)$ and $({1\over 2} R_1 , 0)$ and extending along the vertical direction are conjugates pairs. This implies that their R-R charges are equal and opposite. On the other hand the branes have positive R-R charge, and thus fermions localised at the intersections sitting on the O7 plane are (say) symmetrised while those sitting on the O7 anti-planes (equal in number) are anti-symmetrised. As a result, their net contributions to ${\scr M}$ is zero, consistently with the expression \moeborb .

After $S$ and $P$ modular transformations to the tree-level channel the low-lying untwisted states in $\Kt$, $\At$ and $\Mt$ yield the familiar conditions
$$
\sum_{\alpha =1}^{m_{\rm e} + m_{\rm o}} \, {\bf L}_\alpha \, \left(
N_\alpha + \bar N _\alpha \right) = 32 \left( {\bf R} + {1\over {\bf R}} \right) 
$$
for the NS-NS dilaton tadpole, and 
$$
\sum_{\alpha =1}^{m_{\rm e} + m_{\rm o}} \, {\bf L}_\alpha \, 
\left( N_\alpha \, e^{2i\phi_\alpha \cdot \eta} + \bar N_\alpha \,
e^{-2i\phi_\alpha \cdot \eta} \right) = 32 \, {\bf R}
$$
for the R-R tadpole with, as usual, $\eta^\varLambda$ the chirality of the internal spinors.
In both expressions ${\bf L}_\alpha$ and ${\bf R}$ are defined as in \radii , with $\varLambda =1,2$ running now over the two $T^2$'s.

Much more interesting are the twisted tadpoles since, as repeatedly stated in the previous sub-sections, they clearly display the geometry of the brane configuration. After an $S$ modular transformation one then gets the massless tadpoles
\eqn\twrri{
C_4 O_4\, : \quad \sum_{i=1}^{m_{\rm o}} \, \left( N_i - \bar N_i \right) \, P(z_p \in D_i)  =0 \,, \qquad \forall\, z_p \in \{ x\} \,,
}
and
\eqn\twrrii{
S_4 O_4 \, : \quad \sum_{a=1}^{m_{\rm e}} \, (N_a -\bar N_a ) \, P ( z_q \in D_a) +
\sum_{i=1}^{m_{\rm o}} \, (N_i - \bar N_i ) \, P (z_q \in D_i ) =0 \,, \qquad \forall \, z_q \in \{\xi\}
\,,
}
where
$D_\alpha$ denotes the straight line drawn by the $\alpha$-th brane with wrapping numbers
$(\omega^\varLambda_\alpha , \kappa^\varLambda_\alpha)$, and
$$
P (w_\ell \in D_\alpha ) = 
\cases{
1 & if $w_\ell \in D_\alpha$
\cr
0 & if $w_\ell \not\in D_\alpha$
\cr}
$$
equals one or vanishes depending on whether the point $w_\ell$ belongs to the line $D_\alpha$ or does not. In the simple case where all branes pass through the origin, the 
$P (w_\ell \in D_\alpha ) $ are collected in tables 5 and 6. 

\vskip 10pt

\vbox{\ninepoint
\settabs = 7\columns
\hskip 3.2truecm\vbox{\hrule width  11 truecm}
\+ 
&& $z_p$ & $P (z_p \in D_i) $ 
\cr
\hskip 3.2truecm\vbox{\hrule width  11 truecm}
\+
&& $x_1$ & $\varPi (\kappa^1_i)\, \varPi (\omega^1_i +1)$
\cr
\+
&& $x_2$ & $\varPi (\kappa^1_i)\, \varPi (\omega^1_i +1)\, \varPi (\kappa^2_i) \, \varPi (\omega^2_i +1) $
\cr
\+
&& $x_3$ & $\varPi (\kappa^1_i)\, \varPi (\omega^1_i +1) \, \varPi (\kappa^2_i+1) \, \varPi (\omega^2_i) $
\cr
\+
&& $x_4$ & $\varPi (\kappa^1_i)\, \varPi (\omega^1_i +1) \, \varPi (\kappa^2_i+1) \, \varPi (\omega^2_i +1) $
\cr
\+
&& $x_5$ & $\varPi (\kappa^1_i+1)\, \varPi (\omega^1_i +1)$
\cr
\+
&& $x_6$ & $\varPi (\kappa^1_i+1)\, \varPi (\omega^1_i +1)\, \varPi (\kappa^2_i) \, \varPi (\omega^2_i +1) $
\cr
\+
&& $x_7$ & $\varPi (\kappa^1_i+1)\, \varPi (\omega^1_i +1)\, \varPi (\kappa^2_i+1) \, \varPi (\omega^2_i) $
\cr
\+
&& $x_8$ & $\varPi (\kappa^1_i+1)\, \varPi (\omega^1_i +1)\, \varPi (\kappa^2_i+1) \, \varPi (\omega^2_i +1) $
\cr
\hskip 3.2truecm\vbox{\hrule width  11 truecm}
\cleartabs
\vskip 10pt
\centerline{\ninepoint{\bf Table 5.} Condition for the $\alpha$-th brane to pass through the fixed point $x_p$.}
\vskip 10pt
}

Turning to the twisted NS-NS tadpoles, they can be easily deduced from \twrri\ and \twrrii\ and from their analogy with the untwisted R-R tadpoles. Denoting as usual by $\eta^\varLambda = \pm {1\over 2}$ the chirality of the internal SO(2) spinors, one finds
$$
\eqalign{
O_4 S_4  \,: \quad & \sum_{i=1}^{m_{\rm o}} \, \left( N_i \, e^{2 i\phi_i \cdot \eta}- \bar N_i
\, e^{2 i\phi_i \cdot \eta} \right) \, P(z_p \in D_i)  
\cr
\sim & \sum_{i=1}^{m_{\rm o}} \, N_i \, \sin ( 2 \phi_i \cdot \eta ) \, P(z_p \in D_i) 
\,, \qquad \forall\, z_p \in \{ x\}
\cr}
$$
and, similarly 
$$
O_4 C_4 \,:\quad \sum_{a=1}^{m_{\rm e}} \, N_a \, \sin ( 2\phi_a \cdot \eta)  \, P ( z_q \in D_a) +
\sum_{i=1}^{m_{\rm o}} \, N_i \, \sin ( 2 \phi_i \cdot \eta) \, P (z_q \in D_i )
\,, \quad \forall \, z_q \in \{\xi\} \,.
$$
As expected, in non-supersymmetric models these tadpoles cannot be imposed to vanish, and yield additional contributions to the low-energy effective action \us.

Also in this case, additional non-chiral fermions originating from branes that are parallel in the second $T^2$ can be made massive if a Scherk-Schwarz deformation is acting also
in this torus. In this case, however, one is consequently reshuffling the fixed points in the
two sets $\xi$ and $x$ and thus some care is needed in determining the correct spectrum of the brane intersections.

\vskip 10pt

\vbox{\ninepoint
\settabs = 7\columns
\hskip 3.2truecm\vbox{\hrule width  11 truecm}
\+ 
&& $z_q$ & $P (z_q \in D_i) $ 
\cr
\hskip 3.2truecm\vbox{\hrule width  11 truecm}
\+
&& $\xi_1$ & 1
\cr
\+
&& $\xi_2$ & $\varPi (\kappa^2_\alpha) \, \varPi (\omega^2_\alpha +1) $
\cr
\+
&& $\xi_3$ & $\varPi (\kappa^2_\alpha+1) \, \varPi (\omega^2_\alpha) $
\cr
\+
&& $\xi_4$ & $\varPi (\kappa^2_\alpha+1) \, \varPi (\omega^2_\alpha +1) $
\cr
\+
&& $\xi_5$ & $\varPi (\kappa^1_a+1)\, \varPi (\omega^1_a )$
\cr
\+
&& $\xi_6$ & $\varPi (\kappa^1_a+1)\, \varPi (\omega^1_a)\, \varPi (\kappa^2_a) \, \varPi (\omega^2_a +1) $
\cr
\+
&& $\xi_7$ & $\varPi (\kappa^1_a+1)\, \varPi (\omega^1_a) \, \varPi (\kappa^2_a+1) \, \varPi (\omega^2_a) $
\cr
\+
&& $\xi_8$ & $\varPi (\kappa^1_a+1)\, \varPi (\omega^1_a) \, \varPi (\kappa^2_a+1) \, \varPi (\omega^2_a +1) $
\cr
\hskip 3.2truecm\vbox{\hrule width  11 truecm}
\cleartabs
\vskip 10pt
\centerline{\ninepoint{\bf Table 6.} Condition for the $\alpha$-th brane to pass through the fixed point $\xi_q$.}
\vskip 10pt
}

\subsec{Finale: an interesting example}

To conclude with this deformed $T^4/\bb{Z}_2$ compactification let us present a simple  solution to tadpole conditions. We consider three sets of $N_\alpha$ coincident branes with wrapping numbers
\eqn\wrapnum{
(\omega^\varLambda_1 , \kappa^\varLambda_1) = \left(
\matrix{ (1,0) \cr (1,1)\cr}\right) \,, \quad
(\omega^\varLambda_2 , \kappa^\varLambda_2) = \left(
\matrix{ (2,1) \cr (0,1)\cr}\right) \,, \quad
(\omega^\varLambda_3 , \kappa^\varLambda_3) = \left(
\matrix{ (1,2) \cr (1,-1)\cr}\right) \,, 
}
as depicted in figure 13. These numbers, together with the choice $N_1 = 10$, $N_2 =12$ and $N_3 =6$ clearly satisfy the R-R tadpole conditions. 

\vbox{
\vskip 10pt
%\centerline{\includegraphics[width=2in]{zero-mode.pdf}}
\epsfxsize 10truecm
\centerline{\epsffile{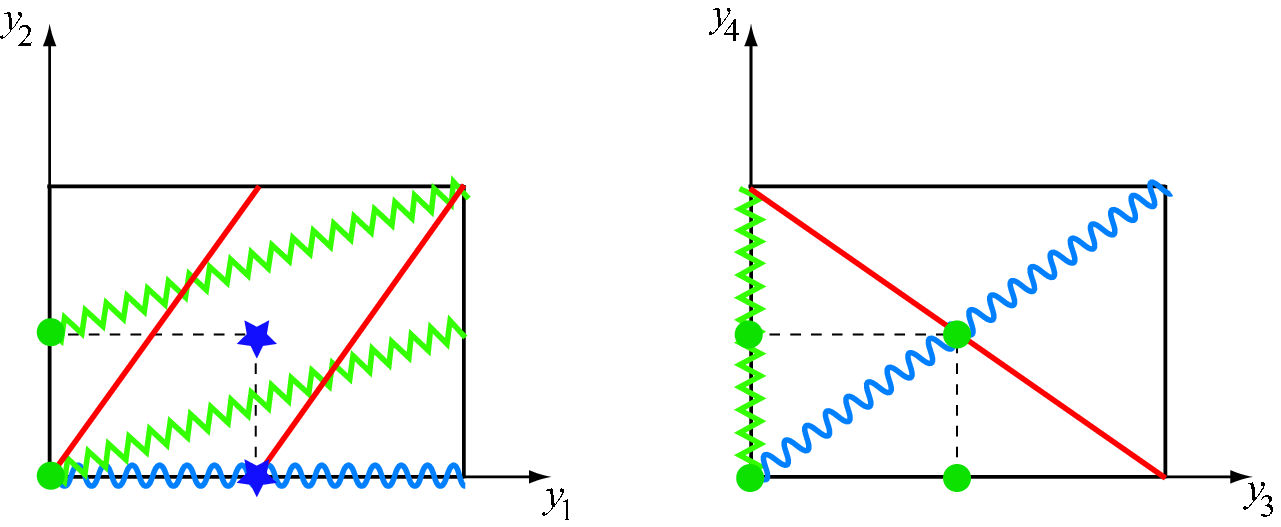}}
\vskip 10pt
\centerline{\ninepoint {\bf Figure 13.} A simple example with three sets of branes.}
\vskip 10pt
}

At the level of toroidal compactification (i.e. before we mod out by the orbifold action and before we deform the spectrum {\it \`a la} Scherk-Schwarz) the chiral spectrum comprises left-handed fermions in the representation $2\, (10,12,1) + 5\, (1,12,6) + 4 \, (1,66,1)$ together with right-handed fermions in the representation $4\, (10,1,6 ) + 3\, (1,12,6) + 8\, (1,1,15) + 2\, (1,66,1) + 2 \, (1,78,1)$, and is clearly free of any irreducible gravitational and gauge anomaly. Before we project and deform the spectrum, let us comment about one subtlety that one meets in deriving this chiral spectrum. In particular, from table 7 one would naively deduce that the $2\, \bar 2$ sector is non-chiral since the corresponding intersection number is zero. However, since $K_2=4\not=0$,  there is a non-vanishing $N_2$-contribution in the M\"obius strip amplitude that apparently is not matched by the annulus. This is evidently not the case and the correct interpretation is as follows: despite the $N_2$ branes and their images are parallel in the second $T^2$ they intersect non-trivially in the first torus. Moreover, the non-chirality of the annulus suggests that left-handed fermions live at four intersections, and similarly do the right-handed ones, while the ``chirality'' of the M\"obius clearly states that only the left-handed fermions lie on top of the orientifold planes. As a result, one gets four copies of left-handed spinors in the antisymmetric representation of ${\rm U}(12)$ together with two copies of right-handed spinors in the symmetric plus antisymmetric representations. 

\vskip 10pt

\vbox{\ninepoint
\settabs = 7\columns
\hskip 3.2truecm\vbox{\hrule width  9 truecm}
\+ 
&& $\alpha\beta$ & $I_{\alpha\beta}$ & $I'_{\alpha\beta}$ & $I'' _{\alpha\beta}$
\cr
\hskip 3.2truecm\vbox{\hrule width  9 truecm}
\+
&& $12$ & $1$ & $1$ & $0$
\cr
\+
&& $13$ & $-4$ & $2$ & $2$
\cr
\+
&& $23$ & $-3$ & $1$ & $0$
\cr
\+
&& $1\bar 2$ & $1$ & $1$ & $0$
\cr
\+
&& $1\bar 3$ & $0$ & $2$ & $2$
\cr
\+
&& $2\bar 3$ & $5$ & $1$ & $0$
\cr
\+
&& $1\bar 1$ & $0$ & $2$ & $0$
\cr
\+
&& $2\bar 2$ & $0$ & $2$ & $0$
\cr
\+
&& $3\bar 3$ & $-8$ & $2$ & $2$
\cr
\hskip 3.2truecm\vbox{\hrule width  9 truecm}
\cleartabs
\vskip 10pt
\centerline{\ninepoint{\bf Table 7.} Intersection numbers for the three sets of branes.}
\vskip 10pt
}

We can now turn on the Scherk-Schwarz deformation and simultaneously project the spectrum by the geometrical $\bb{Z}_2$. From \wrapnum\ one can deduce that fermions in the dipole sector of the $N_1$ and $N_3$ branes will get a mass proportional to the compactification radius, while the neutral sector of the $N_2$ branes stays supersymmetric. Moreover, the branes intersect at different fixed points, and in particular some intersections of the $N_1$ and $N_3$ branes coincide with fixed points in both sets $\{\xi_q\}$ and $\{x_p\}$. Using the explicit value of the intersection numbers in table 7, together with the general expressions in the previous sub-section one gets the following chiral massless spectrum: right-handed fermions in the adjoint representation of the U(12) gauge group and in the representations $2\, (10,1,6) + 2 \, (1,12,6) +4\, (1,1,15) + (1,66,1) + (1,78,1)$, together with left-handed fermions in the representations $(10,12,1) + 3\, (1,12,6) + 3\, (1,66,1) + (1,78,1)$. As usual, the cancellation of R-R tadpoles guarantees that the spectrum is free of irreducible gravitational and gauge anomalies, while the Green-Schwarz mechanism takes care of the reducible anomaly \refs{\greeni,\greenii}
$$
{\scr I}_{8} = {\textstyle{1\over 4}} \left( {\rm tr}\, F_2^2 - 2 {\rm tr}\, F_3^2 \right) \left( 
{\rm tr}\, F_1^2 + {\rm tr}\, F_3^2 - {\textstyle{1\over 2}} {\rm tr}\, R^2 \right) \,.
$$

\vskip 24pt
\noindent
{\bf Acknowledgement} We would like to thank Ignatios Antoniadis, Ralph Blumenhagen, Emilian Dudas and Amine Hammou for useful discussions. C.A. would like to thank the Theory Unit at CERN and the Physics Department in Heraklion for hospitality. M.C. would like to thank the Physics Department of the Ludwig-Maximilians University and the Theory Unit at CERN for hospitality. N.I. would like to thank the Physics Department of the Humboldt University in Berlin and the Theory Unit at  CERN for hospitality. The work of C.A. was supported by the Alexander von Humboldt Stiftung. The work of M.C. was partially supported by INFN and MURST. This work was supported in part by the European Community's Human Potential Programme under the contract HPRN-CT-2004-005104 to which both the Ludwig-Maximilians University, the University of Turin and the University of Milan 1 belong, and in part by the European Community's Human Potential Programme under the contract HPRN-CT-2004512194, to which the University of Crete belongs.

\listrefs

\end